\newcommand{\MSbar}{\overline{\rm MS}}
\newcommand{\DRbar}{\overline{\rm DR}}
\newcommand{\DRbarprime}{\overline{\rm DR}'}

\newcommand{\zetathree}{\zeta_3}
\newcommand{\zetafive}{\zeta_5}
\newcommand{\afour}{{\rm Li}_4(1/2)}
\newcommand\beq{\begin{eqnarray}}
\newcommand\eeq{\end{eqnarray}}

\newcommand{\Pitilde}{\widetilde \Pi}
\newcommand{\Pibar}{\overline \Pi}
\newcommand{\myhat}{\widehat}
\newcommand{\hath}{\widehat h}
\newcommand{\Mgpole}{M_{\tilde g}}
\newcommand{\mgrun}{m_{\tilde g}}
\newcommand{\mghat}{\widehat m_{\tilde g}}
\newcommand{\CG}{C_A}
\newcommand{\Cq}{C_q}
\newcommand{\Iq}{I_q}
\newcommand{\Lg}{\ell_{\tilde g}}
\newcommand{\Lsq}{\ell_{\tilde Q}}
\newcommand{\LG}{L_{\tilde g}}
\newcommand{\LSQ}{L_{\tilde Q}}
\newcommand{\LR}{L}
\newcommand{\Lr}{\ell}

\newcommand{\AaO}{a^{(1)}_0}
\newcommand{\Aaa}{a^{(1)}_1}
\newcommand{\AbO}{a^{(2)}_0}
\newcommand{\Aba}{a^{(2)}_1}
\newcommand{\Abb}{a^{(2)}_2}
\newcommand{\AcO}{a^{(3)}_0}
\newcommand{\Aca}{a^{(3)}_1}
\newcommand{\Acb}{a^{(3)}_2}
\newcommand{\Acc}{a^{(3)}_3}

\newcommand{\AaOO}{a^{(1)}_{00}}
\newcommand{\AaaO}{a^{(1)}_{10}}
\newcommand{\AaOa}{a^{(1)}_{01}}
\newcommand{\AbOO}{a^{(2)}_{00}}
\newcommand{\AbaO}{a^{(2)}_{10}}
\newcommand{\AbOa}{a^{(2)}_{01}}
\newcommand{\AbbO}{a^{(2)}_{20}}
\newcommand{\Abaa}{a^{(2)}_{11}}
\newcommand{\AbOb}{a^{(2)}_{02}}
\newcommand{\AcOO}{a^{(3)}_{00}}
\newcommand{\AcaO}{a^{(3)}_{10}}
\newcommand{\AcOa}{a^{(3)}_{01}}
\newcommand{\AcbO}{a^{(3)}_{20}}
\newcommand{\Acaa}{a^{(3)}_{11}}
\newcommand{\AcOb}{a^{(3)}_{02}}
\newcommand{\AccO}{a^{(3)}_{30}}
\newcommand{\Acba}{a^{(3)}_{21}}
\newcommand{\Acab}{a^{(3)}_{12}}
\newcommand{\AcOc}{a^{(3)}_{03}}

\newcommand{\AbOOp}{\overline a^{(2)}_{00}}
\newcommand{\AbaOp}{\overline a^{(2)}_{10}}
\newcommand{\AbOap}{\overline a^{(2)}_{01}}
\newcommand{\AcOOp}{\overline a^{(3)}_{00}}
\newcommand{\AcaOp}{\overline a^{(3)}_{10}}
\newcommand{\AcOap}{\overline a^{(3)}_{01}}
\newcommand{\AcbOp}{\overline a^{(3)}_{20}}
\newcommand{\Acaap}{\overline a^{(3)}_{11}}
\newcommand{\AcObp}{\overline a^{(3)}_{02}}

\newcommand{\CMc}{c^{(3)}_{\mgrun}}

\newcommand{\Bhata}{b^{(1)}_{\myhat h}}
\newcommand{\Bhatb}{b^{(2)}_{\myhat h}}
\newcommand{\Bhatc}{b^{(3)}_{\myhat h}}
\newcommand{\BMhata}{b^{(1)}_{\mghat}}
\newcommand{\BMhatb}{b^{(2)}_{\mghat}}
\newcommand{\BMhatc}{b^{(3)}_{\mghat}}

\newcommand{\Ba}{b^{(1)}_{h}}
\newcommand{\Bb}{b^{(2)}_{h}}
\newcommand{\Bc}{b^{(3)}_{h}}
\newcommand{\Bn}{b^{(n)}_{h}}
\newcommand{\BMa}{b^{(1)}_{\mgrun}}
\newcommand{\BMb}{b^{(2)}_{\mgrun}}
\newcommand{\BMc}{b^{(3)}_{\mgrun}}
\newcommand{\BMn}{b^{(n)}_{\mgrun}}
\newcommand{\BQa}{b^{(1)}_{m_{\tilde Q}^2}}
\newcommand{\BQba}{b^{(2a)}_{m_{\tilde Q}^2}}
\newcommand{\BQbb}{b^{(2b)}_{m_{\tilde Q}^2}}

\documentclass[twocolumn,
amsmath,
prd,nofootinbib,floatfix,
]{revtex4}

\allowdisplaybreaks
\usepackage{graphicx}

\begin{document}
\renewcommand{\theequation}{\arabic{section}.\arabic{equation}}

\title{Refined gluino and squark pole masses beyond leading order}

\author{Stephen P. Martin}
\affiliation{
Physics Department, Northern Illinois University, DeKalb IL 60115 USA\\
{\rm and}
Fermi National Accelerator Laboratory, PO Box 500, Batavia IL 60510}


\begin{abstract} 
The physical pole and running masses of squarks and gluinos have recently 
been related at two-loop order in a mass-independent renormalization 
scheme. I propose a general method for improvement of such formulas, and 
argue that better accuracy results. The improved version gives an 
imaginary part of the pole mass that agrees exactly with the direct 
calculation of the physical width at next-to-leading order. I also find 
the leading three-loop contributions to the gluino pole mass in the case 
that squarks are heavier, using effective field theory and renormalization 
group methods. The efficacy of these improvements for the gluino and 
squarks is illustrated with numerical examples. Some necessary three-loop 
results for gauge coupling and fermion mass beta functions and pole masses 
in theories with more than one type of fermion representation, which are 
not directly accessible from the published literature, are presented in an 
Appendix.
\end{abstract}

\maketitle


\tableofcontents

\section{I\lowercase{ntroduction}}\label{sec:introduction}
\setcounter{equation}{0}
\setcounter{footnote}{1}

The small ratio of the electroweak symmetry breaking scale to the Planck 
mass can be stabilized \cite{quadscancel} in softly-broken supersymmetric 
extensions of the Standard Model. This implies that all of the Standard 
Model particles will have superpartners, which should be within reach of 
the $p\overline p$ Fermilab Tevatron collider or the $pp$ Large Hadron 
Collider during the next few years. Most of the new parameters appearing 
in the Minimal Supersymmetric Standard Model (MSSM) \cite{Martin:1997ns} 
are the masses of the new superpartners and other supersymmetry-breaking 
couplings of positive mass dimension. Therefore, a detailed understanding 
of the MSSM Lagrangian is nearly synonymous with an understanding of 
supersymmetry breaking.

The fact that experimental observations of flavor violation and CP 
violation are not in significant disagreement with the predictions of the 
Standard Model can be taken as indirect evidence for the existence of some
powerful organizing principle governing supersymmetry breaking and its 
mediation to the MSSM sector. An especially interesting possibility is 
that the organizing principle can be discerned by running the parameters 
of the theory up to high energy scales using the renormalization group. To 
carry out this analysis, it will be crucial to relate physically measured 
observables, especially the superpartner masses, to running parameters in 
the full theory defined by the non-decoupled Lagrangian that includes all 
of the superpartners.

However, running masses are not the most direct observables expected from 
collider experiments. In general, the mass defined by the position of the 
complex pole in the propagator is a gauge-invariant and renormalization 
scale-invariant quantity \cite{Tarrach:1980up}-\cite{Gambino:1999ai}. The 
pole mass does suffer from ambiguities \cite{poleambiguities} due to 
infrared physics associated with the QCD confinement scale, but these are 
probably not large enough to cause a practical problem for 
strongly-interacting superpartners.  The complex pole mass should be 
closely related in a calculable way to the kinematic observable mass and 
width reported by experiments \cite{massdefs}.

It is often useful to calculate in on-shell schemes, in which some 
physical masses and other observables are used as input data and others 
are outputs. However, for the key purpose of unraveling the organizing 
principle behind the supersymmetry-breaking Lagrangian, this is not as 
directly useful. The $\MSbar$ scheme \cite{MSbar} can also be used, but it 
violates supersymmetry explicitly. Instead, it is preferable to use the 
$\DRbar$ scheme \cite{DRbar} (or the revised $\DRbarprime$ scheme 
\cite{DRbarprime}, which removes the unphysical effects of epsilon-scalar 
masses in softly-broken supersymmetric models), with all superpartners 
non-decoupled. While it is difficult to know in detail what limitations on 
this program will follow from future experimental uncertainties, it seems 
clear that multi-loop calculations will be necessary to make the 
theoretical sources of error negligible.

The one-loop relations between the superpartner pole masses and the 
running parameters in the MSSM Lagrangian have been known for some time 
\cite{Martin:1993yx}-\cite{PBMZ}. The calculation of the Higgs scalar 
boson masses in the MSSM has now advanced to include the important 
two-loop corrections (for some reviews of recent progress, see 
\cite{Degrassi:2002fi}-\cite{Heinemeyer:2004ms}), and even some three-loop 
corrections \cite{Degrassi:2002fi}, using a variety of different methods. 
Recent calculations have provided the supersymmetric QCD (SUSYQCD) 
two-loop corrections to the squark \cite{Martin:2005eg} and gluino 
\cite{Yamada:2005ua,Martin:2005ch} masses. The quark masses in the MSSM 
are known at two-loop order \cite{quarkpoleSUSY},\cite{Martin:2005ch}. 
More generally, refs.~\cite{Martin:2005eg} and \cite{Martin:2005ch} 
provide the self-energy functions and pole masses for scalars and 
fermions, respectively, calculated in mass-independent renormalization 
schemes at two-loop order in any renormalizable field theory, in the 
approximation that vector bosons are treated as massless in the two-loop 
parts. This approximation is likely to be quite good for most applications 
to the MSSM, because the largest two-loop effects involving vector bosons 
come from SUSYQCD, and because the $W$ and $Z$ bosons are evidently 
lighter than most of the superpartners.

It is important to consider the validity of the perturbative expansion in 
these results. Especially for the lightest Higgs boson, the squarks, and 
the gluino, the corrections that give the pole masses from the running 
masses turn out to be quite significant. As a prominent example, even the 
pure two-loop correction to the gluino mass (compared to the running 
mass evaluated at a renormalization group scale equal to itself) is of 
order 1-2\% in the case that squarks and gluinos are comparable in mass, 
and grows to about 5\% for squark masses that are of order 5 times heavier 
than the gluino. One would like some assurance that perturbation theory is 
really converging, and an estimate of the theoretical error. 
Unfortunately, the renormalization-scale dependence of these results is 
not a reliable error estimate; in particular, the scale dependence of 
one-loop corrections is routinely much smaller than the two-loop 
corrections when the latter are known.

Another reason to be wary is the fact that calculations in 
mass-independent renormalization schemes like $\MSbar$ or $\DRbarprime$ 
use propagators with masses that can differ significantly from the 
physical ones. In many cases this is true for any reasonable choice of the 
renormalization scale $Q$. A troubling aspect of this is that the 
imaginary part of the complex pole squared mass,
\beq
s_{\rm pole} = M^2 - i \Gamma M
\label{eq:definespole}
\eeq
can give a numerical value for the width $\Gamma$ that differs quite badly 
from the physical width. It is not hard to find examples for which the 
tree-level masses are sufficiently different from the physical masses that 
a particular contribution to $\Gamma$ as computed from the complex pole 
mass is exactly 0 (because the decay would be kinematically forbidden if 
the particles had masses equal to the tree-level Lagrangian masses 
appearing in the propagators of the self-energy loop diagrams), while the 
true decay width contribution (computed directly from diagrams with 
multi-particle final states, using an on-shell scheme) is non-zero. Or, 
the reverse can happen. (I will show an example of each type in Figure 
\ref{fig:widths} of section \ref{sec:numerical}.) While the complex pole 
mass is in principle a gauge-invariant and renormalization-scale invariant 
observable, this calls into question how well one can trust the 
perturbation theory that yields it in practice.

These issues are general. However, in the MSSM, they are particularly 
acute for the squarks and the gluino, because of their strong coupling. 
Furthermore, the LHC will quite likely produce gluinos and squarks in 
abundance if supersymmetry is correct. Therefore, I will use the squark 
and gluino SUSYQCD system within the MSSM as an example in this paper to 
show how to ameliorate the problems mentioned above. First, in section 2, 
I discuss how to reorganize the results of perturbation theory by 
expanding tree-level masses around physical masses in the loop corrections 
obtained in mass-independent ($\MSbar$ or $\DRbarprime$) schemes. In 
section 3, I present a result for the three-loop corrections to the gluino 
mass, valid in the limit that squarks are treated as nearly degenerate and 
much heavier than the gluino. This is the case where one might expect 
three-loop and even higher-order corrections to be most dangerous, but I 
show that they are actually under good control, and can be tamed by using 
effective field theory and renormalization group methods. Section 4 
displays some numerical results showing the efficacy of these 
improvements. In an Appendix, I present some necessary results for 
three-loop contributions to fermion mass beta functions and pole masses in 
(non-supersymmetric) theories with fermions in distinct representations.

\section{I\lowercase{mproved two-loop pole mass results}} 
\setcounter{equation}{0} 
\setcounter{footnote}{1}

In general, the two-loop order expression for the pole mass of a particle
can be computed from knowledge of the self-energy function, 
\beq
\Pi_{j}^k(s) = \frac{1}{16\pi^2} \Pi^{(1)k}_{j} 
+  \frac{1}{(16\pi^2)^2} \Pi^{(2)k}_{j} + \ldots,
\eeq
Here $s$ is the external momentum invariant, the superscript in 
parentheses indicates the loop order, and the indices $j,k$ indicate 
different particles with the same quantum numbers, which in general can 
mix. For fermions, $\Pi_{j}^{k}(s)$ can be assembled from separate 
chirality-preserving and chirality-violating self-energy functions, as 
described in section II.C of ref.~\cite{Martin:2005ch}. Then the 
gauge-invariant and renormalization-scale invariant pole squared masses 
can be defined formally as the solutions to the equation
\beq
{\rm Det}[(s - m_j^2) \delta_{j}^{k} - \Pi_{j}^{k}(s)] = 0,
\eeq
where $m_i^2$ are the tree-level diagonalized squared masses. However, 
because $\Pi_{j}^{k}$ should be interpreted as a complex-valued function 
of a real variable $s$, this equation must be solved by first expanding 
the self-energy function as a series about a point on the real $s$-axis. 
In evaluating the loop integrals in the self-energy, $s$ is given a {\em 
positive} infinitesimal imaginary part, while the complex pole squared 
mass solution [see eq.~(\ref{eq:definespole})] always has a {\em 
non-positive} imaginary part. A related subtlety is that when the particle 
in question has couplings to massless gauge bosons, terms of a given loop 
order in the self-energy have branch-cut singularities (except when the 
Fried-Yennie gauge-fixing condition is used 
\cite{PasseraSirlin,Martin:2005eg}).

The most straightforward way to obtain the pole mass at two-loop order 
in a mass-independent renormalization scheme is 
to first expand $\Pi_{j}^k(s)$ in a series about the tree-level squared 
masses. Define, for a generic squared mass $m^2$:
\begin{widetext}
\beq
\Pitilde_{j}^{(1)k}(m^2) &\equiv& \lim_{s \rightarrow m^2 + i \varepsilon}
  \Pi^{(1)k}_{j}(s)
  ,
\\
\Pitilde_{j}^{(2)j}(m^2) &\equiv& \lim_{s \rightarrow m^2 + i \varepsilon}
  \left [ \Pi^{(2)j}_{j}(s) 
+ \Pi_{j}^{(1)j}(s) \frac{\partial}{\partial s}\Pi_{j}^{(1)j}(s) \right ]
,
\label{eq:defPitildetwo}
\eeq
where the self-energy functions on the right-hand side are computed in a 
mass-independent renormalization scheme, and no sum on $j$ is implied in 
eq.~(\ref{eq:defPitildetwo}). Then, working consistently to two-loop 
order, the pole mass for the particle with tree-level squared mass $m^2_j$ 
is
\beq 
M_j^2 - i \Gamma_j M_j 
&=& 
m_j^2 
+ \frac{1}{16 \pi^2} \Pitilde_{j}^{(1)j}(m_j^2) 
+ \frac{1}{(16 \pi^2)^2} \Bigl \lbrace \Pitilde_{j}^{(2)j}(m_j^2) 
+ \sum_{k\not= j} 
\Pitilde_{j}^{(1)k}(m_j^2) \Pitilde_{k}^{(1)j}(m_j^2)/(m_j^2 - m_k^2) 
\Bigr \rbrace ,
\label{eq:mpolegen}
\eeq
obtained by a perturbative\footnote{Here the loop-induced mixing between 
particles $j$ and $k$ is assumed small compared to the tree-level squared 
mass splitting. Otherwise, one must perform almost-degenerate perturbation 
theory, by expanding around a modified tree-level Lagrangian designed to 
minimize the one-loop mixing in that sector. This could plausibly occur 
for Higgsino-like neutralinos in the MSSM.} expansion of 
eq.~(\ref{eq:definespole}).

The previous expression is gauge-invariant, and renormalization scale 
invariant up to terms of three-loop order. In its application to the 
squark and gluino masses in the MSSM, this approach has the advantage of 
depending only on tree-level running parameters, so that iteration is not 
necessary if they are taken as given. However, as remarked in the 
Introduction, the use of tree-level running masses in propagators is 
problematic at least for the imaginary part of the pole mass, which arises 
from the absorptive part of the self-energy functions. If the tree-level 
masses differ significantly from the physical masses of the particles, 
then the kinematics of the self-energy functions will poorly reflect the 
actual kinematics giving rise to the physical width of the particle. This 
can lead to a non-zero width when there should be none, or vice versa. In 
general, one may care more about the real part of the pole mass, but 
intuitively one cannot expect the real part to be very accurate if the 
kinematics in the loop integrations poorly reflects the physical particle 
masses, and if the imaginary part is completely wrong.

To improve the situation, let us reorganize the previous result by 
expanding all tree-level squared masses appearing in the $\Pitilde$ 
functions in a series about the real parts of their respective pole 
squared masses. (Note that although I have written the $\Pitilde$ 
functions as depending on a single external squared mass argument that 
replaced $s$, there is also dependence on the internal propagator masses 
which is not explicitly indicated.) Doing this, one arrives at:
\beq 
M_j^2 - i \Gamma_j M_j 
&=& 
m_j^2 
+ \frac{1}{16 \pi^2} \Pibar_{j}^{(1)j}(M_j^2) 
+ \frac{1}{(16 \pi^2)^2} \Bigl \lbrace \Pibar_{j}^{(2)j}(M_j^2) 
+ \sum_{k\not= j}  
\Pibar_{j}^{(1)k}(M_j^2) 
\Pibar_{k}^{(1)j}(M_j^2)/(M_j^2 - M_k^2) 
\nonumber \\ && 
- \sum_k {\rm Re}\bigl [\Pibar_{k}^{(1)k}(M_k^2) \bigr ] 
\frac{\partial}{\partial M_k^2} \Pibar_{j}^{(1)j}(M_j^2) 
\Bigr \rbrace 
,
\label{eq:mpolegenimp}
\eeq
The second sum over $k$ is taken over all fermions and bosons in the 
theory that couple to particle $j$, not just those that can mix with $j$. 
The $\Pibar$ in eq.~(\ref{eq:mpolegenimp}) are defined to have the same 
functional dependence on the real parts of the pole squared masses (both 
external and internal) as the functions $\Pitilde$ in 
eq.~(\ref{eq:mpolegen}) did on the tree-level squared masses. Because the 
one-loop and two-loop parts of eq.~(\ref{eq:mpolegen}) are each separately 
gauge-fixing invariant, it is clear that eq.~(\ref{eq:mpolegenimp}) is 
also independent of gauge-fixing. Formally, eqs.~(\ref{eq:mpolegen}) and 
(\ref{eq:mpolegenimp}) are equivalent up to terms of three-loop order. 
However, in eq.~(\ref{eq:mpolegenimp}), all kinematic dependences of loop 
integrals on the right-hand side correspond to the physical masses (the 
real parts of the pole masses). I therefore expect that, when it makes 
much of a difference, eq.~(\ref{eq:mpolegenimp}) should be more accurate 
than eq.~(\ref{eq:mpolegen}). If one starts with the Lagrangian parameters 
as input, evaluation of the pole masses will require an iterative 
procedure involving all of the particle masses simultaneously, which in a 
general case at two-loop order could take a significant computation time. 
On the other hand, if physical masses are taken as inputs, 
eq.~(\ref{eq:mpolegenimp}) still requires knowledge of the tree-level 
couplings and mixing matrices of the theory. The one-loop functions are 
always known analytically in terms of logarithms, so taking derivatives of 
them poses no technical difficulties. The two-loop functions often cannot 
be computed analytically, but computer codes such as {\tt TSIL} 
\cite{TSIL} provide for their numerical computation.

To illustrate the method, I choose here to consider the two-loop 
corrections to the gluino and squark pole masses, for simplicity including 
only SUSYQCD corrections and ignoring the effects of squark mixing, quark 
masses, and electroweak effects. For this system, the realization of 
eq.~(\ref{eq:mpolegen}) is:
\beq 
&&
\!\!\!\!\!\!
\!\!\!\!\!\!
M_{\tilde Q_j}^2 - i \Gamma_{\tilde Q_j} M_{\tilde Q_j} 
\>=\> m_{\tilde Q_j}^2 
+ h \Cq \Pitilde_{\tilde Q}^{(1)}(m_{\tilde Q_j}^2, \mgrun^2)   
+ h^2 \Cq
 \Bigl [
 \Cq \Pitilde_{\tilde Q}^{(2,a)}(m_{\tilde Q_j}^2, \mgrun^2) 
+
 \CG \Pitilde_{\tilde Q}^{(2,b)}(m_{\tilde Q_j}^2, \mgrun^2) 
\nonumber \\ &&
\qquad\qquad\qquad\quad
+  \Iq \sum_k \Pitilde_{\tilde Q}^{(2,c)}
   (m_{\tilde Q_j}^2, \mgrun^2, m_{\tilde Q_k}^2 ) 
\Bigr ]
,
\label{eq:mpolesquarksnonimp}
\\
&&
\!\!\!\!\!\!
\!\!\!\!\!\!
\Mgpole^2 - i \Gamma_{\tilde g} \Mgpole 
\>=\> 
\mgrun^2 + h
 \Bigl [\CG \Pitilde_{\tilde g}^{(1,a)}(\mgrun^2) +
 \Iq \sum_j \Pitilde_{\tilde g}^{(1,b)}(\mgrun^2, m_{\tilde Q_j}^2) 
 \Bigr ]
+ h^2 \Bigl [
 \CG^2 \Pitilde_{\tilde g}^{(2,a)}(\mgrun^2) 
\nonumber \\ && 
\qquad\qquad\quad
 +\CG \Iq \sum_j \Pitilde_{\tilde g}^{(2,b)}(\mgrun^2, m_{\tilde Q_j}^2) 
 +\Cq \Iq \sum_j \Pitilde_{\tilde g}^{(2,c)}(\mgrun^2, m_{\tilde Q_j}^2) 
 +\Iq^2 \sum_j \sum_k \Pitilde_{\tilde g}^{(2,d)}
  (\mgrun^2, m_{\tilde Q_j}^2, m_{\tilde Q_k}^2) 
\Bigr ]
.
\label{eq:mpolegluinononimp}
\eeq
Here, $m_{\tilde g}^2$ and $m_{\tilde Q_j}^2$ are the tree-level 
$\DRbarprime$ squared masses. In many references, $m_{\tilde g}$ is 
written as $M_3$, but in the present paper, capital letters are reserved 
for pole masses and lowercase letters for running masses. The strong gauge 
coupling appears in the combination:
\beq
h \>\equiv\> 
{g_3^2}/{16 \pi^2} \>=\> \alpha_S/4\pi.
\eeq
The indices $j$ and $k$ run over the 12 squark mass eigenstates of the 
MSSM (taken here to be unmixed, but not necessarily degenerate), and for 
$SU(3)_c$, $\CG = 3$ and $\Cq = 4/3$ and $\Iq = 1/2$. The functions 
$\Pitilde$ appearing in eq.~(\ref{eq:mpolesquarksnonimp}) were given in 
eqs.~(5.6)-(5.9) of ref.~\cite{Martin:2005eg}, and those in 
eq.~(\ref{eq:mpolegluinononimp}) were given in eqs.~(5.5)-(5.10) of 
ref.~\cite{Martin:2005ch}. Note that the dependence on all squared masses 
is explicit in the arguments of these functions.

Applying eq.~(\ref{eq:mpolegenimp}) to this gives the improved 
equations for the relation between pole and running squared masses:
\beq
&&
\!\!\!\!\!\!
\!\!\!\!\!\!
M_{\tilde Q_j}^2 - i \Gamma_{\tilde Q_j} M_{\tilde Q_j} 
\>=\> m_{\tilde Q_j}^2 
+ h \Cq \Pitilde_{\tilde Q}^{(1)}(M_{\tilde Q_j}^2, M_{\tilde g}^2)   
+ h^2 \Cq
 \Bigl \lbrace
 \Cq \Pitilde_{\tilde Q}^{(2,a)}(M_{\tilde Q_j}^2, M_{\tilde g}^2) 
+
 \CG \Pitilde_{\tilde Q}^{(2,b)}(M_{\tilde Q_j}^2, M_{\tilde g}^2) 
\nonumber \\ &&
\qquad\qquad\qquad\>\>\>
+  \Iq \sum_k \Pitilde_{\tilde Q}^{(2,c)}(M_{\tilde Q_j}^2, 
M_{\tilde g}^2, M_{\tilde Q_k}^2 )
- \Cq {\rm Re}[\Pitilde_{\tilde Q}^{(1)}(M_{\tilde Q_j}^2, M_{\tilde g}^2)]  
\frac{\partial}{\partial M_{\tilde Q_j}^2} 
\Pitilde_{\tilde Q}^{(1)}(M_{\tilde Q_j}^2, M_{\tilde g}^2)     
\nonumber \\ &&
\qquad\qquad\qquad\>\>\>
- {\rm Re}\bigl [\CG \Pitilde_{\tilde g}^{(1,a)}(M_{\tilde g}^2) +
 \Iq \sum_k \Pitilde_{\tilde g}^{(1,b)}(M_{\tilde g}^2, M_{\tilde Q_k}^2) 
\bigr ] 
\frac{\partial}{\partial M_{\tilde g}^2} 
\Pitilde_{\tilde Q}^{(1)}(M_{\tilde Q_j}^2, M_{\tilde g}^2)     
\Bigr \rbrace,
\label{eq:mpolesquarksimp}
\\
&&
\!\!\!\!\!\!
\!\!\!\!\!\!
\Mgpole^2 - i \Gamma_{\tilde g} \Mgpole 
\>=\> \mgrun^2 
+ h 
 \Bigl [\CG \Pitilde_{\tilde g}^{(1,a)}(M_{\tilde g}^2) +
 \Iq \sum_j \Pitilde_{\tilde g}^{(1,b)}(M_{\tilde g}^2, M_{\tilde Q_j}^2) 
 \Bigr ]
+ h^2 
 \biggl \lbrace
 \CG^2 \Pitilde_{\tilde g}^{(2,a)}(M_{\tilde g}^2) 
\nonumber \\ && 
\qquad\qquad\quad
+
 \CG \Iq \sum_j \Pitilde_{\tilde g}^{(2,b)}(M_{\tilde g}^2, M_{\tilde Q_j}^2) 
 +
 \Cq \Iq \sum_j \Pitilde_{\tilde g}^{(2,c)}(M_{\tilde g}^2, M_{\tilde Q_j}^2) 
 +
 \Iq^2 \sum_j \sum_k \Pitilde_{\tilde g}^{(2,d)}
 (M_{\tilde g}^2, M_{\tilde Q_j}^2, M_{\tilde Q_k}^2) 
\nonumber \\ &&
\qquad\qquad\quad
- {\rm Re} 
\Bigl [\CG \Pitilde_{\tilde g}^{(1,a)}(M_{\tilde g}^2) +
 \Iq \sum_k \Pitilde_{\tilde g}^{(1,b)}(M_{\tilde g}^2, M_{\tilde Q_k}^2) 
 \Bigr ]
\frac{\partial}{\partial M_{\tilde g}^2} 
\Bigl [\CG \Pitilde_{\tilde g}^{(1,a)}(M_{\tilde g}^2) +
 \Iq \sum_j \Pitilde_{\tilde g}^{(1,b)}(M_{\tilde g}^2, M_{\tilde Q_j}^2) 
 \Bigr ]
\nonumber \\ &&
\qquad\qquad\quad
- \Cq \Iq \sum_j {\rm Re} 
\Bigl [\Pitilde_{\tilde Q}^{(1)}(M_{\tilde Q_j}^2, M_{\tilde g}^2) \Bigr ]
\frac{\partial}{\partial M_{\tilde Q_j}^2} 
\Pitilde_{\tilde g}^{(1,b)}(M_{\tilde g}^2, M_{\tilde Q_j}^2) 
\biggr \rbrace .
\label{eq:mpolegluinoimp}
\eeq
Given the running masses and coupling as inputs, the pole masses can now 
be solved for iteratively. Or, given the pole masses and the 
running gauge coupling as inputs, the running masses can be immediately 
extracted.

I have checked analytically that the imaginary parts of 
the pole masses given in eqs.~(\ref{eq:mpolesquarksimp}) and 
(\ref{eq:mpolegluinoimp}) correspond exactly to the gluino and squark 
decay widths calculated at next-to-leading order in 
\cite{Beenakker:1996dw}. This is a good reason to prefer the improved 
version eqs.~(\ref{eq:mpolesquarksimp}) and (\ref{eq:mpolegluinoimp}) 
over eqs.~(\ref{eq:mpolesquarksnonimp}) and (\ref{eq:mpolegluinononimp}), 
and more generally eq.~(\ref{eq:mpolegenimp}) over 
eq.~(\ref{eq:mpolegen}). In section \ref{sec:numerical}, I will present a 
numerical comparison of these equations.
\end{widetext}
\section{T\lowercase{hree-loop contributions to the gluino pole 
mass}\label{sec:threeloopgluino}}
\setcounter{equation}{0}
\setcounter{footnote}{1}

Radiative corrections to the mass of the gluino in the MSSM are 
particularly large, for two reasons. First, the gluino is a color octet, 
and so is effectively more strongly coupled than a quark in the 
fundamental representation. Second, it couples to 12 squark/quark pairs. 
If one expresses the gluino pole mass in terms of the running mass 
evaluated at itself in a non-decoupling scheme, then the squark-mediated 
corrections are large and grow logarithmically with the ratio of the 
squark to gluino masses. One can exploit this by using effective field 
theory and renormalization group methods to obtain the 
logarithmically-enhanced parts. In this section, I will use this strategy 
to evaluate the three-loop gluino pole mass in the formal limit $M_{\tilde 
Q}^2 \gg M_{\tilde g}^2$, neglecting terms of three-loop order that are 
suppressed by $M_{\tilde g}^2/M_{\tilde Q}^2$, but including all terms of 
order $L^3$, $L^2$, and $L$, where
\beq
\LR \>\equiv\> \ln(M_{\tilde Q}^2/M_{\tilde g}^2).
\eeq
I will also find the coefficient of the $L^0$ term, up to a single 
(presently) unknown and plausibly sub-dominant 
matching coefficient. This analysis neglects squark 
mixing and non-degeneracy, electroweak effects, and Standard 
Model quark masses, for simplicity. The same method could quite easily be 
extended to include all terms of order
\beq
\alpha_S^n \LR^{n}, \>\>
\alpha_S^n \LR^{n-1}, \>\>{\rm and}\>\>
\alpha_S^n \LR^{n-2}, 
\eeq
at arbitrary loop order $n$, but the residual unknown three-loop order 
contributions are likely to be larger than the contributions from $n\geq 
4$. In the following, I will use the following convenient notations for 
other logarithms:
\beq
\LG &\equiv& \ln (\Mgpole^2/Q^2),
\\
\LSQ &\equiv& \ln (M_{\tilde Q}^2/Q^2),
\\
\Lr &\equiv& \ln(m_{\tilde Q}^2/m_{\tilde g}^2),
\\
\Lg &\equiv& \ln (m_{\tilde g}^2/Q^2),
\\
\Lsq &\equiv& \ln (m_{\tilde Q}^2/Q^2),
\eeq
where $Q$ is the renormalization group scale, and the running gluino and 
squark masses in the last three definitions are taken to be in the full 
theory (with squarks included).

The starting point is the three-loop pole mass for a color octet Majorana 
fermion ($\tilde g$) in the presence of 6 light quark flavors 
$(u,d,s,c,b,t)$. This is the same fermion content of SUSYQCD as 
found in ref.~\cite{slightlysplitsusy} and the extreme limit of
``split supersymmetry" \cite{splitsusy}; it is the 
effective theory in which squarks have been decoupled. The three-loop 
gluino pole mass in this model is almost known from 
ref.~\cite{Melnikov:2000qh}, up to a single ambiguity that is resolved in 
the Appendix of the present paper. The result can be written as:
\begin{widetext}
\beq
\Mgpole &=&
\mghat(Q) \Bigl ( 1 
+ \hath \bigl [\Aaa \LG + \AaO\bigr ]
+ \hath^2 \bigl [\Abb \LG^2 + \Aba \LG + \AbO\bigr ]
+ \hath^3 \bigl [
    \Acc \LG^3 + \Acb \LG^2 + \Aca \LG + \AcO \bigr ]
+ \ldots \Bigr ),
\phantom{xxx.}
\label{eq:MgpoledecA}
\eeq
\end{widetext}
where $\Mgpole$ is the gluino pole mass, and the hats on the symbols 
$\mghat$ and $\hath \equiv \hat g_3^2/16\pi^2$ are used to distinguish the 
running $\MSbar$ parameters in the effective theory without squarks. The 
coefficients appearing here are:
\beq
\AaO &=& 12,
\qquad 
\Aaa \>=\> -9,
\\
\AbO &=& 339 + 33\pi^2 - 36 \pi^2 \ln 2 + 54 \zetathree
,
\\ 
\Aba &=& -282,
\qquad
\Abb \>=\> 63,
\\ 
\AcO &=& 
172399/18 
+ 73367 \pi^2/10 
- 12672 \pi^2 \ln 2 
\nonumber \\ &&
+ 12330 \zetathree 
- 121 \pi^4/2 
- 48 \pi^2 \ln^2 2 
\nonumber \\ &&
+ 552 \ln^4 2 
+ 13248 \afour 
\nonumber \\ &&
+ 675 \pi^2 \zetathree 
- 1890 \zetafive
,
\\ 
\Aca &=& 
  -8932 - 627 \pi^2 + 684 \pi^2\ln2 - 306 \zetathree
,
\\ 
\Acb &=& 3093 ,
\qquad
\Acc \>=\> -399.
\eeq
Note that in this section, the gluino pole mass is real, because the 
gluino has no allowed decays in pure SUSYQCD in the case that all squarks 
are heavier.

The terms in eq.~(\ref{eq:MgpoledecA}) that depend on $\LG$, and thus 
explicitly involve the renormalization scale $Q$, are obtained from the 
renormalization group equations for the gauge coupling and running gluino 
mass in the effective theory:
\beq
Q\frac{d}{dQ} \hath &=& 
  \hath^2 \Bhata + \hath^3 \Bhatb + \hath^4 \Bhatc +\ldots 
\label{eq:betahath}
\\
Q\frac{d}{dQ} \myhat \mgrun &=& \myhat \mgrun \bigr [
  \hath \BMhata + \hath^2 \BMhatb + \hath^3 \BMhatc +\ldots \bigr ],
\phantom{xzx}
\eeq
where
\beq
\Bhata &=& -10,
\qquad
\Bhatb \>=\> 44, 
\\
\Bhatc &=& 4168/3, 
\\
\BMhata &=& -18,
\qquad
\BMhatb \>=\> -228, 
\\ 
\BMhatc &=& -2774 + 1440 \zetathree.
\eeq
The last coefficient does not seem to be obtainable directly from the 
results in the published literature, which only deal with theories with a 
single type of fermion representation. However, it can be inferred from an 
unpublished paper of O.V.~Tarasov \cite{Tarasov:1982gk}. This is explained 
in the Appendix of the present paper.

The results given above are not what is needed in order to discern an 
organizing principle for supersymmetry breaking. Instead, one needs to 
obtain the running parameters in the full theory with squarks included. 
This can be achieved from matching conditions between the effective theory 
parameters $(\hath, \mghat)$ and the full theory parameters $(h, \mgrun, 
m_{\tilde Q}^2)$. In these equations, the effective theory parameters 
$(\hath, \mghat)$ are always in $\MSbar$, while the full theory parameters 
$(h, \mgrun, m_{\tilde Q}^2)$ are always in the $\DRbarprime$ scheme here. 
(It may also be possible to use $\DRbar$ for the effective theory, but 
then there are subtleties associated with evanescent couplings in 
non-supersymmetric theories \cite{nonSUSYDRbar}.) Now one can take the 
gluino pole mass from eq.~(\ref{eq:MgpoledecA}) and use the matching to 
rewrite it in terms of the parameters of the full theory.  
In the approximation used here, non-renormalizable terms 
are not included in the effective field theory Lagrangian, which 
corresponds to neglecting all contributions suppressed by powers of 
$M_{\tilde g}^2/M_{\tilde Q}^2$. Experience shows that expansions of 
radiative corrections in such mass ratios typically converge quite 
quickly, so I suspect that it is reasonable to hope that the results below 
will be approximately valid even if the typical squark mass is not very 
much larger than the gluino mass.

The renormalization group equations for the parameters of the
non-decoupled theory are:
\beq
Q\frac{d}{dQ} h &=& h^2 \Ba + h^3 \Bb + h^3 \Bc + \ldots 
\label{eq:betahthree}
\\
Q\frac{d}{dQ} \mgrun &=& 
  \mgrun \bigr [h \BMa + h^2 \BMb + h^3 \BMc + \ldots \bigr ]
  \phantom{xxx}
\\
Q\frac{d}{dQ} m^2_{\widetilde Q} &=& 
  \mgrun^2 \bigl [h \BQa + h^2 \BQba + \ldots \bigr ] 
\nonumber \\ && 
+ m^2_{\widetilde Q} \bigl [h^2 \BQbb + \ldots \bigr ]
\eeq
where \cite{threeloopgaugeMSSM}:
\beq
\Ba &=& -6,
\qquad
\Bb \>=\> 28, 
\qquad
\Bc \>=\> 694/3, \phantom{\frac{X'^X}{X'_X}}
\eeq
and \cite{twoloopgauginoMSSM,threeloopgauginoMSSM}:
\beq
\BMn &=& n \Bn,
\eeq
and \cite{DRbarprime,twoloopsquarkMSSM}:
\beq
\BQa &=& -32/3,
\qquad 
\BQba \>=\> -128/3,\phantom{X_{X_X}}
\\ 
\BQbb &=& 64.\phantom{\frac{X^X}{X^X}}
\label{eq:endfullRGE}
\eeq

The two-loop gauge coupling matching condition has been obtained in 
ref.~\cite{Harlander:2005wm}:
\begin{widetext}
\beq
\myhat h &=& h \Bigl [1 
+ h \bigl (-1 + 2 \LSQ \bigr )
+ h^2 \bigl (23 - 12 \LSQ + 4 \LSQ^2 \bigr ) + \ldots \Bigr ] .
\label{eq:matchh}
\eeq
For the gaugino mass matching condition, I find:
\beq 
\mghat &=& \mgrun \Bigl [1 
+ h \bigl (6 \LSQ \bigr ) 
+ h^2 \bigl (14 \pi^2/3 - 176 + 133 \LSQ + 18 \LSQ^2 \bigr )
\nonumber \\ &&
+ h^3 \bigl (
\CMc + [1713 - 720 \zetathree] \LSQ + 615\LSQ^2  + 48 \LSQ^3
\bigr ) + \ldots \Bigr ].
\label{eq:matchMg}
\eeq
I obtained the two-loop coefficients here by a direct comparison of the 
two-loop gluino pole squared mass as found in the full theory with 
$\DRbarprime$ and in the effective theory with $\MSbar$, using the results 
of ref.~\cite{Martin:2005ch}. The three-loop logarithmic terms in 
eq.~(\ref{eq:matchMg}) are obtained from the renormalization group 
equations (\ref{eq:betahath})-(\ref{eq:endfullRGE}). Unfortunately, the 
coefficient $\CMc$ remains unknown, and seems quite difficult to 
calculate.

The preceding results can now be used straightforwardly to obtain the 
three-loop result for the gluino pole mass in terms of the non-decoupled 
theory parameters. I find:
\beq
\Mgpole &=&
\mgrun(Q) \Bigl ( 1 
+ h \bigr [\AaOO + \AaaO \LG + \AaOa \LSQ \bigl ]
+ h^2 \bigr [\AbOO 
     + \AbaO \LG 
     + \AbOa \LSQ
     + \AbbO \LG^2 
     + \Abaa \LG \LSQ
     + \AbOb \LSQ^2\bigl ]
\nonumber \\ &&
\!\!\!\!\!\!
\!\!\!\!\!\!\!\!
+ h^3 \bigr [\AcOO 
     + \AcaO \LG 
     + \AcOa \LSQ
     + \AcbO \LG^2 
     + \Acaa \LG \LSQ
     + \AcOb \LSQ^2
     + \AccO \LG^3 
     + \Acba \LG^2 \LSQ
     + \Acab \LG \LSQ^2
     + \AcOc \LSQ^3) \bigl ]
\Bigr ),
\phantom{xxxx}
\label{eq:Mgpolenondecimp}
\eeq 
where
\beq 
\AaOO &=& 12, \qquad 
\AaaO \>=\> -9, \qquad
\AaOa \>=\> 6, \qquad
\AbOO \>=\> 151 + 113 \pi^2/3 - 36 \pi^2 \ln 2 + 54\zetathree 
,
\\ 
\AbaO &=& -273, \qquad
\AbOa \>=\> 229, \qquad 
\AbbO \>=\> 63, \qquad
\Abaa \>=\> -72, \qquad
\AbOb \>=\> 18,
 \\
\AcOO &=& 
 \CMc
 + 127147/18 + 73267 \pi^2/10 
 - 12600 \pi^2 \ln 2
 + 12222 \zetathree 
 - 121 \pi^4/2 
 - 48 \pi^2 \ln^2 2 
\nonumber \\ &&
 + 552 \ln^4 2 
 + 13248 \afour  
 + 675 \pi^2 \zetathree 
 - 1890 \zetafive 
,
 \\
\AcOa &=& 6483 + 330 \pi^2  - 360 \pi^2 \ln2   - 180 \zetathree
,\phantom{xx}
\qquad\quad
\AcaO \>=\> -6991 -669 \pi^2  + 684 \pi^2 \ln 2 - 306 \zetathree
,
\\
\AcbO &=& 2967, \qquad
\Acaa \>=\> -3855, \qquad
\AcOb \>=\> 1023, \qquad
\AccO \>=\> -399, \qquad
\\
\Acba &=& 630, \qquad
\Acab \>=\> -306, \qquad
\AcOc \>=\> 48.
\label{eq:endMgpolenondecimp}
\eeq
It is also possible to rewrite this result so that the logarithms involve 
running masses, by using the two-loop relation between the squark running 
and pole masses in the formal limit $M^2_{\tilde Q} \gg M^2_{\tilde g}$, 
obtained from ref.~\cite{Martin:2005eg} using 
eq.~(\ref{eq:mpolesquarksimp}) of the present paper:
\beq
M^2_{\tilde Q} - i \Gamma_{\tilde Q} M_{\tilde Q} = 
m^2_{\tilde Q} +  M^2_{\tilde Q} \Bigl [
h \frac{8}{3}(1 - i \pi)
+ h^2 \Bigl \lbrace \frac{92}{9} + \frac{4 \pi^2}{9} (1 + 8 \ln 2) 
+ 24 \LSQ -
i \pi \Bigl (\frac{932}{9} - \frac{128 \pi^2}{27} - 8 \LSQ \Bigr ) 
\Bigr \rbrace
\Bigr ].
\eeq
The result, formally equivalent to eq.~(\ref{eq:Mgpolenondecimp}) up to
terms of four-loop order, is:
\beq
\Mgpole &=&
\mgrun(Q) \Bigl ( 1 
+ h \bigr [\AaOO + \AaaO \Lg + \AaOa \Lsq \bigl ]
+ h^2 \bigr [\AbOOp 
     + \AbaOp \Lg 
     + \AbOap \Lsq
     + \AbbO \Lg^2 
     + \Abaa \Lg \Lsq
     + \AbOb \Lsq^2\bigl ]
\nonumber \\ &&
\!\!\!\!\!\!
\!\!\!\!\!
+ h^3 \bigr [\AcOOp 
     + \AcaOp \Lg 
     + \AcOap \Lsq
     + \AcbOp \Lg^2 
     + \Acaap \Lg \Lsq
     + \AcObp \Lsq^2
     + \AccO \Lg^3 
     + \Acba \Lg^2 \Lsq
     + \Acab \Lg \Lsq^2
     + \AcOc \Lsq^3)\bigl ]
\Bigr ),
\phantom{xx}
\label{eq:Mgpolenondecnonimp}
\eeq 
where the new coefficients are:
\beq 
\AbOOp &=& -49 + 113 \pi^2/3 - 36 \pi^2 \ln 2 + 54\zetathree 
,\qquad
\qquad
\AbaOp \>=\> -111, \qquad
\qquad
\AbOap \>=\> 121, \qquad 
\\
\AcOOp &=& 
 \CMc +
 60895/18 + 199541 \pi^2/30 
 - (35792/3) \pi^2 \ln 2
 + 11250 \zetathree 
 - 121 \pi^4/2 
 - 48 \pi^2 \ln^2 2 
\nonumber \\ &&
 + 552 \ln^4 2 
 + 13248 \afour  
 + 675 \pi^2 \zetathree 
 - 1890 \zetafive 
,
 \\
\AcaOp &=& 809 - 669 \pi^2 + 684 \pi^2 \ln 2 - 306 \zetathree 
,
\qquad\qquad\quad
\AcOap \>=\> 837 + 330 \pi^2  -360 \pi^2 \ln 2 - 180 \zetathree
,
\\
\AcbOp &=& 294, \qquad
\Acaap \>=\> -723, \qquad
\AcObp \>=\> 159. \qquad
\label{eq:endofMimp}
\label{eq:endMgpolenondecnonimp}
\eeq
For practical calculations, it is useful to extract from the above 
expressions the contributions
that can be consistently added to the complete two-loop results 
of eqs.~(\ref{eq:mpolegluinononimp}) and (\ref{eq:mpolegluinoimp}), or 
more generally eq.~(\ref{eq:mpolegen}) or (\ref{eq:mpolegenimp}). From 
eq.~(\ref{eq:Mgpolenondecnonimp}), I obtain:
\beq
\Delta^{(3)}_{\mbox{eq.~(\ref{eq:mpolegen}) or
(\ref{eq:mpolegluinononimp})}} \Mgpole^2 &=& 
h^3 \mgrun^2 \bigl [ 
   d_{00} 
   + d_{10} \Lr 
   + d_{20} \Lr^2  
   + d_{30} \Lr^3 
   + d_{01} \Lg
   + d_{11} \Lr \Lg 
   + d_{21} \Lr^2 \Lg 
\nonumber \\ &&
   + d_{02} \Lg^2
   + d_{12} \Lr \Lg^2 
   + d_{03} \Lg^3 
\bigr ]
\label{eq:gluinothreenonimp}
\eeq
with coefficients:
\beq
d_{00}  &=& 2 \CMc 
  + 50311/9 
  + 213101 \pi^2/15 
  - (74176/3) \pi^2 \ln 2 
  + 23796 \zetathree
  - 121 \pi^4 
  - 96 \pi^2 \ln^2 2 
\nonumber \\ &&
  + 1104 \ln^4 2 
  + 26496 \afour
  + 1350 \pi^2 \zetathree 
  - 3780 \zetafive
,
\\
d_{10}  &=& 
  3990 + 1112 \pi^2  - 1152 \pi^2 \ln 2 + 288 \zetathree
,
\qquad\qquad
d_{20} \>=\> 2202
,
\qquad\qquad
d_{30} \>=\> 312
,
\\
d_{01} &=& 3826 - 904 \pi^2  + 864 \pi^2 \ln 2 - 1296 \zetathree
,
\qquad\qquad
d_{11} \>=\> -2280,
\qquad\qquad
d_{21} \>=\>-864,
\\
d_{02} &=& -384,  
\qquad\qquad
d_{12} \>=\> 648,
\qquad\qquad
d_{03} \>=\> -108 , 
\eeq
and from 
eq.~(\ref{eq:Mgpolenondecimp}):
\beq
\Delta^{(3)}_{\mbox{eq.~(\ref{eq:mpolegenimp})
or (\ref{eq:mpolegluinoimp})}} \Mgpole^2 &=& 
  h^3 \Mgpole^2 \bigl [ 
   e_{00} 
   + e_{10} \LR 
   + e_{20} \LR^2  
   + e_{30} \LR^3 
   + e_{01} \LG 
   + e_{11} \LR \LG 
   + e_{02} \LG^2 
 \bigr ],\phantom{xx}
\label{eq:gluinothreeimp}
\eeq
where
\beq
e_{00} &=& 2 \CMc 
 + 91507/9 
 + 59707\pi^2/5 
 - 22608 \pi^2 \ln 2
 + 20556 \zetathree 
 + 1104 \ln^4 2 
 - 96 \pi^2 \ln^2 2 
\nonumber \\ &&
 - 121 \pi^4 
 + 26496 \afour 
 + 1350 \pi^2 \zetathree 
 - 3780 \zetafive
,
\\
e_{10} &=& 1410 - 696 \pi^2 + 576 \pi^2 \ln 2 - 2304 \zetathree
,
\\
e_{20} &=& -2310
,
\qquad\quad
e_{30} \>=\> 312, 
\qquad\quad
e_{01} \>=\> -314,
\qquad\quad
e_{11} \>=\> -504,
\qquad\quad
e_{02} \>=\> 126.
\eeq 
[The coefficients of $\LR^2 \LG$ and $\LR \LG^2$ and $\LG^3$ in 
eq.~(\ref{eq:gluinothreeimp}) vanish.] 
Writing these in a convenient numerical form, I find:
\beq
\Delta^{(3)}_{\mbox{eq.~(\ref{eq:mpolegen}) or
(\ref{eq:mpolegluinononimp})}} \Mgpole^2 &=& 
\alpha_S^3 \mgrun^2 \bigl [ 
 0.00101 \CMc 
 + 9.616 
 + 
   3.744 \Lr 
 + 
   1.110 \Lr^2 
 + 
   0.157 \Lr^3 
 - 
   0.375 \Lg 
\nonumber \\ && \quad\qquad
 - 
   1.149 \Lr \Lg 
 - 
   0.435 \Lr^2 \Lg 
 - 
   0.194 \Lg^2 
 + 
   0.327 \Lr \Lg^2 
 - 
   0.0544\Lg^3
\bigr ],
\label{eq:gluinothreenonimpnum}
\\
\Delta^{(3)}_{\mbox{eq.~(\ref{eq:mpolegenimp})
or (\ref{eq:mpolegluinoimp})}} \Mgpole^2 &=& 
  \alpha_S^3 \Mgpole^2 \bigl [ 
 0.00101 \CMc 
 + 5.992
 - 
   2.161 \LR 
 - 
   1.164\LR^2 
 + 
   0.157\LR^3 
\nonumber \\ && \quad\qquad
 + 
   0.158\LG 
 - 
   0.254 \LR \LG 
 + 
   0.0635 \LG^2 
 \bigr ].\phantom{xx}
\label{eq:gluinothreeimpnum}
\eeq
When applied to realistic models, the two-loop gluino pole mass should be 
calculated including all relevant effects including squark mixing and 
Yukawa and electroweak couplings using eq.~(\ref{eq:mpolegen}) or 
(\ref{eq:mpolegenimp}), and then the appropriate corresponding formula 
eq.~(\ref{eq:gluinothreenonimp}) or (\ref{eq:gluinothreeimp}) can be added 
with an approximate overall squark mass scale parameterized by either 
$\Lr$ or $\LR$. (However, the three-loop contributions found here must be 
eschewed if most of the squarks are not heavier than the gluino.)

The parts of 
eqs.~(\ref{eq:gluinothreenonimpnum})-(\ref{eq:gluinothreeimpnum}) 
that do not contain logarithms came from two sources: the 
non-logarithmic part of eq.~(\ref{eq:MgpoledecA}), and the unknown 
three-loop gluino mass matching coefficient $\CMc$ defined in 
eq.~(\ref{eq:matchMg}). It is useful to note that the dominant part of the 
contribution from eq.~(\ref{eq:MgpoledecA}) is due to loop diagrams 
containing only gluons and light quark internal lines. Furthermore, there 
is a significant partial cancellation in this non-logarithmic piece, not 
due to supersymmetry, which is not present in the effective theory, but to 
the accident of the number of quarks in the Standard Model. To see this, 
one can tag the contributions according to the number $n$ of closed gluino 
loops in each diagram, and also treat the number of light quarks $n_Q$ 
(equal to 6 in the real world) as a variable,
using eq.~(\ref{eq:MvRplus}). Then, in 
eqs.~(\ref{eq:gluinothreenonimpnum}) and (\ref{eq:gluinothreeimpnum})
respectively:
\beq
 9.616 &\rightarrow&
33.312 - 26.634 (n_Q/6) + 3.489 (n_Q/6)^2 - 0.027 (n_Q/6)^3
+ [-0.598 - 0.0095 (n_Q/6)] n + 0.084 n^2,\phantom{xx}
\\
 5.992 &\rightarrow& 16.703 - 10.543 (n_Q/6) + 0.667 (n_Q/6)^2 + 0.027 
(n_Q/6)^3 + [-1.072 + 0.126 (n_Q/6)] n + 0.084 n^2 .
\eeq 
Thus the diagrams with no closed heavy particle loops dominate the final 
result, and would even more so if it were not for the tendency of gluon 
and quark loops to cancel. In the numerical studies of the following 
section, I will simply neglect the effects of the unknown coefficient 
$\CMc$, since it comes from diagrams with at least one heavy squark loop 
and therefore is plausibly less significant than the other non-logarithmic 
contributions. Note that $|\CMc|$ would have to exceed $10^4$ in order for 
it to contribute 1\% to the gluino pole mass formula.

It is also useful to observe that there will typically be a significant 
cancellation between the logarithmic and non-logarithmic contributions in 
eq.~(\ref{eq:gluinothreeimpnum}). Therefore, that version of the 
three-loop contribution to the pole mass is actually considerably smaller 
than one might have naively suspected.


\section{N\lowercase{umerical results}\label{sec:numerical}}
\setcounter{equation}{0}
\setcounter{footnote}{1}

For purposes of illustration, consider a simplified model, with all 
squarks degenerate and unmixed, and quark masses and electroweak effects 
neglected, as in the previous section. The pertinent Lagrangian parameters 
are then the running SUSYQCD coupling $\alpha_S(Q)$ and the gluino and 
common squark masses in the $\DRbarprime$ scheme. Throughout this section, 
I will fix $\alpha_S(M_{\tilde g}) = 0.095$.

In the left panel of Figure \ref{fig:pole}, I compare different 
computations of the ratio of the real part of the gluino pole mass 
to the running mass 
evaluated at the pole mass, $M_{\tilde g}/m_{\tilde g}(M_{\tilde g})$.%
\begin{figure}[tbp]  
\centering
\mbox{\includegraphics[width=8.8cm]{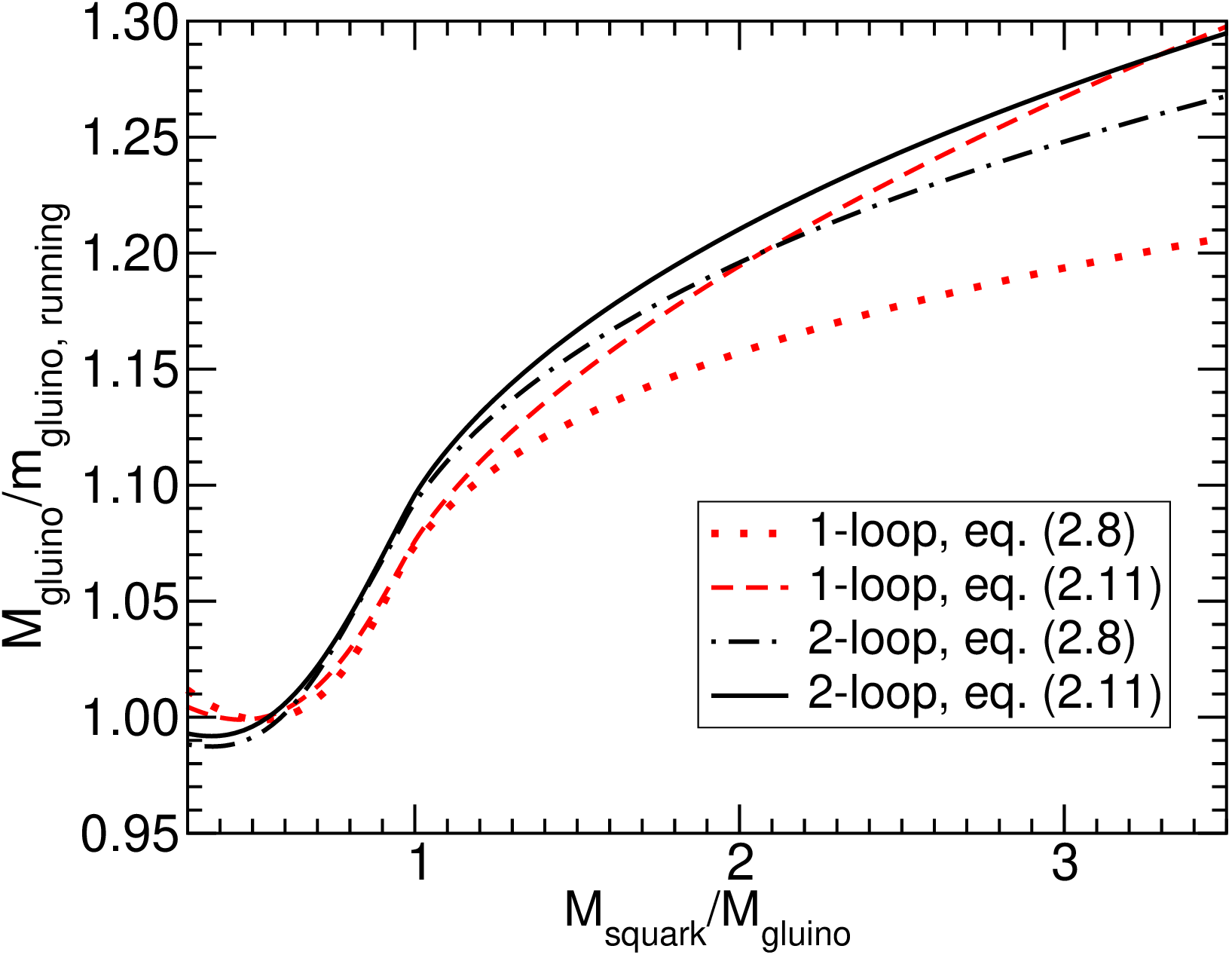}
~
\includegraphics[width=8.8cm]{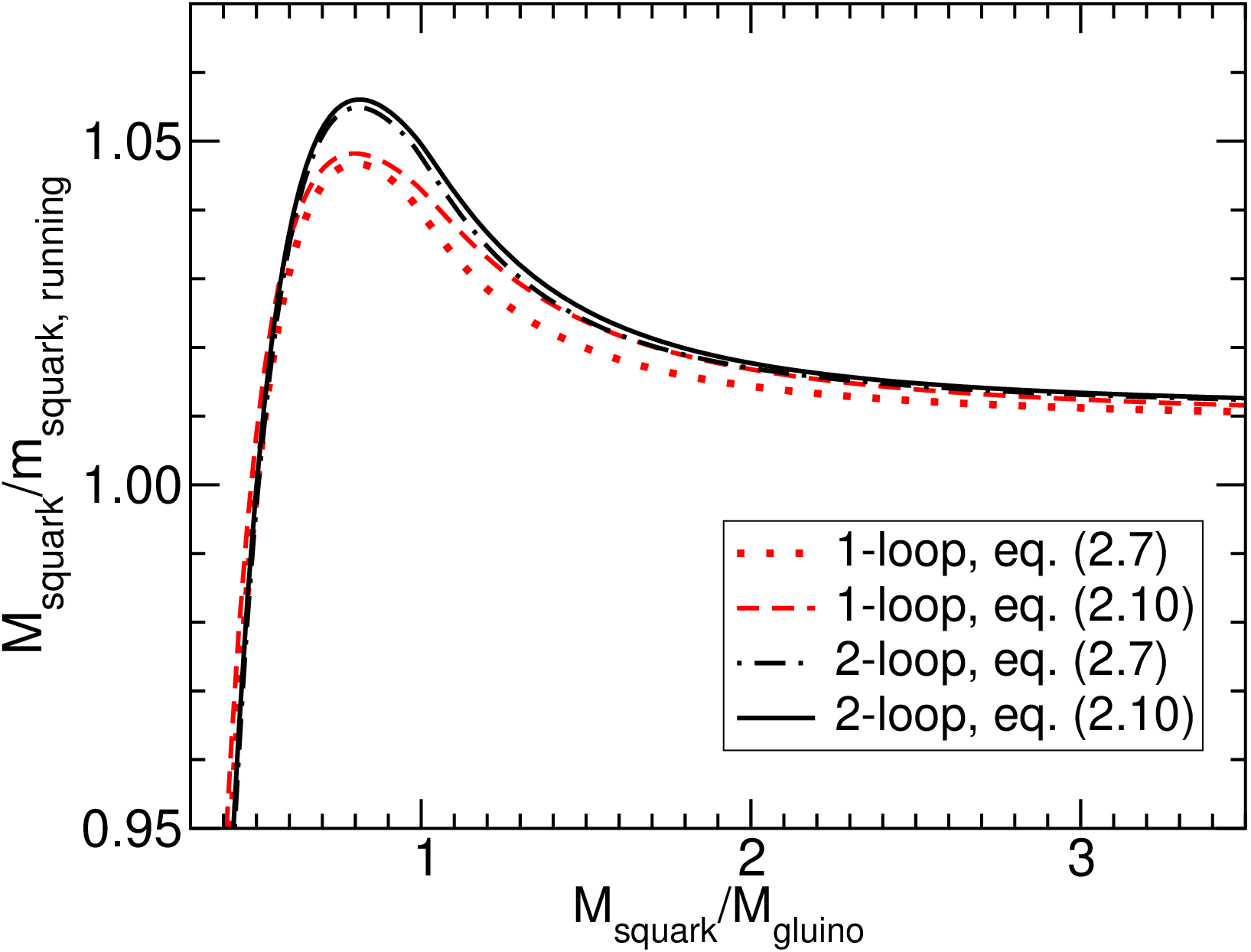}}
\caption{\label{fig:pole}
The ratios of the gluino and squark pole masses to the running masses, 
$M_{\tilde g}/m_{\tilde g}(M_{\tilde g})$ in the left panel and $M_{\tilde 
Q}/m_{\tilde Q}(M_{\tilde Q})$ in the right panel, as functions of the 
ratio of the common squark pole mass to the gluino pole mass, $M_{\tilde 
Q}/M_{\tilde g}$. The different lines correspond to the two-loop results 
of eqs.~(\ref{eq:mpolegluinononimp}) and (\ref{eq:mpolegluinononimp}) and 
the one-loop truncations of the same formulas. For simplicity, here all 
squarks are taken to be degenerate and unmixed, and quark masses and 
electroweak effects are neglected. The renormalization scale use for the 
computation is $Q=M_{\tilde g}$ for the left panel and $Q=M_{\tilde Q}$ in 
the right panel, and the SUSYQCD coupling is fixed to $\alpha_S(M_{\tilde 
g}) = 0.095$ in both cases.}
\end{figure}
The two-loop computations of eqs.~(\ref{eq:mpolegluinononimp}) and 
(\ref{eq:mpolegluinoimp}) for the pole mass agree to better than 1\% for 
$M_{\tilde Q}/M_{\tilde g} < 1.55$, but the disagreement increases for 
larger values of that ratio, and reaches 2.7\% when $M_{\tilde 
Q}/M_{\tilde g} = 3.5$. It is in just this regime that the three-loop 
contributions to the gluino pole mass found above should be reliable and 
important; I will return to this below.

In the right panel of Figure \ref{fig:pole}, the same comparison is done 
for the ratio of the real part of the
squark pole mass to the running mass. Here, the 
different two-loop computations are in extremely close agreement over the 
entire range. Furthermore, the overall magnitude of the radiative 
corrections is much smaller than for the gluino. 
I conclude that purely theoretical 
uncertainties for squark masses are probably under control at a level of 
much better than 1 per cent. (The steep ``cliff" at the left side of the 
graph reflects the fact that a much heavier gluino makes a large negative 
radiative contribution to the squark pole squared mass.)

I expect that the solid lines in Figure \ref{fig:pole}, reflecting the 
calculations of eqs.~(\ref{eq:mpolesquarksimp}) and 
(\ref{eq:mpolegluinoimp}), are more reliable than those of 
eqs.~(\ref{eq:mpolesquarksnonimp}) and (\ref{eq:mpolegluinononimp}). As 
discussed above, one reason for this expectation is the fact that the 
former equations do a much better job of approximating the decay widths 
(the imaginary parts of the pole mass) in the near-threshold region. This 
is illustrated in Figure \ref{fig:widths}.%
\begin{figure}[tpb]  
\centering
\mbox{\includegraphics[width=8.8cm]{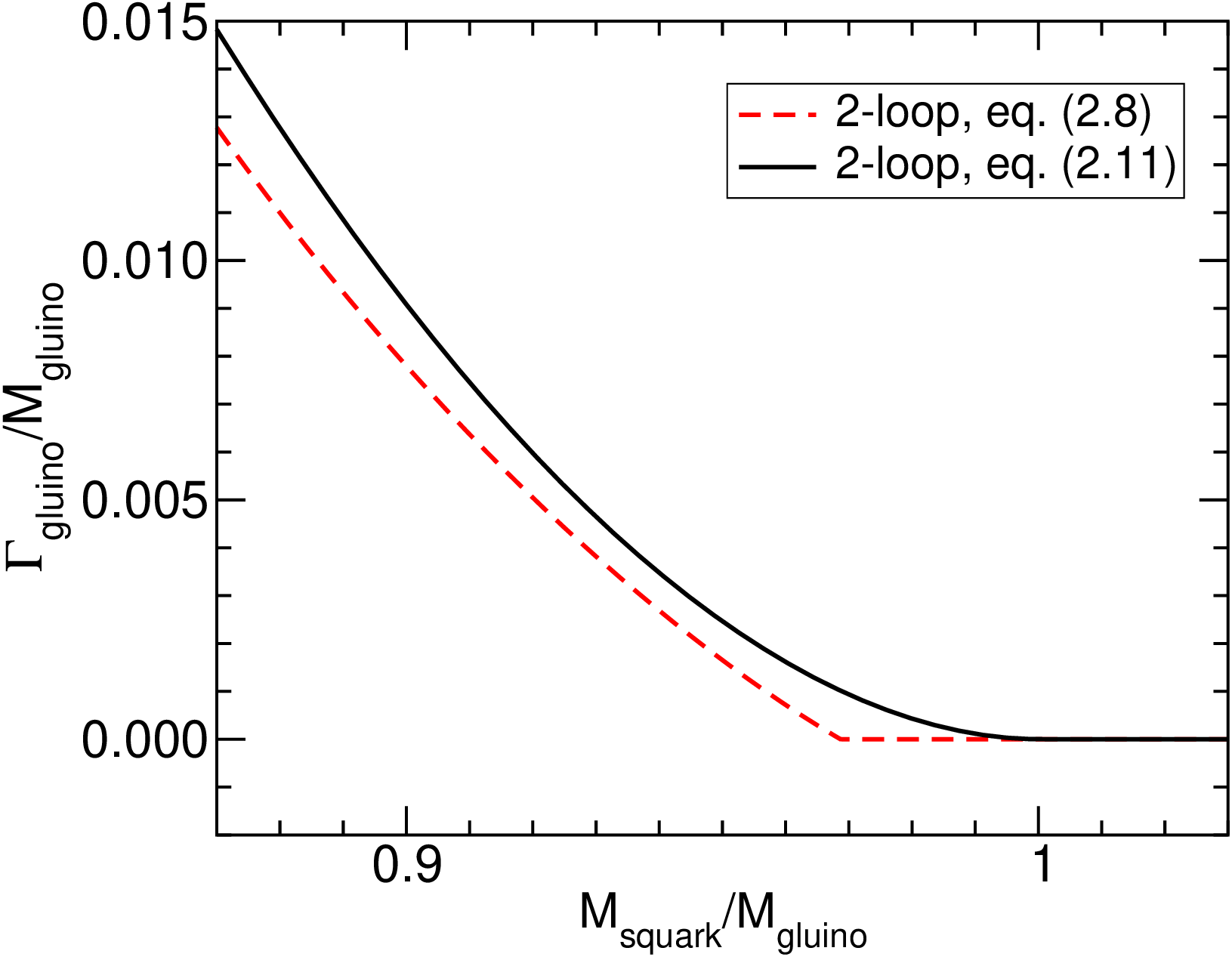}
~
\includegraphics[width=8.8cm]{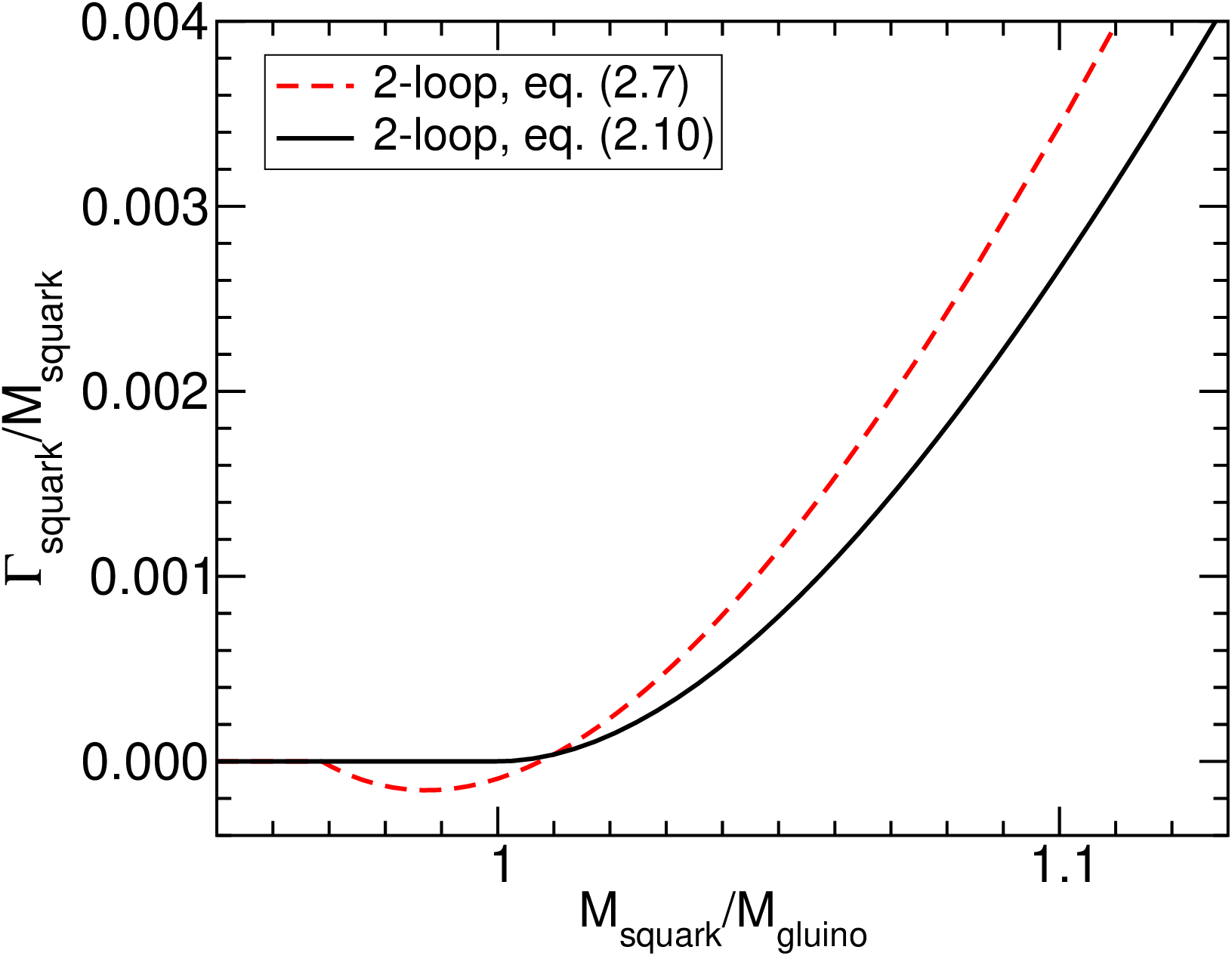}}
\caption{\label{fig:widths}
The gluino and squark widths as extracted from the complex pole masses, 
$\Gamma_{\tilde g}/M_{\tilde g}$ in the left panel and $\Gamma_{\tilde 
Q}/M_{\tilde Q}$ in the right panel, as functions of the ratio of the 
squark pole mass to the gluino pole mass, in the near-threshold region 
of parameter space. The different lines correspond to different two-loop 
approximations, as defined in the text. The widths as computed using 
eqs.~(\ref{eq:mpolesquarksimp}), (\ref{eq:mpolegluinoimp}) agree exactly 
with the direct next-to-leading order width calculation of 
ref.~\cite{Beenakker:1996dw}, while the widths as computed using 
eqs.~(\ref{eq:mpolesquarksnonimp}), (\ref{eq:mpolegluinononimp}) fail to 
agree with the decay kinematics dictated by the physical masses near 
threshold.}
\end{figure}
First, the gluino width $\Gamma_{\tilde g}$ as calculated from the pole 
mass using eq.~(\ref{eq:mpolegluinononimp}) actually vanishes for all 
$M_{\tilde Q} > 0.969 M_{\tilde g}$, rather than for $M_{\tilde Q} \geq 
M_{\tilde g}$ as dictated by kinematics. The reason for this failure is 
that the width in the pole mass derives from the imaginary parts of loop 
integrals which, in that approximation, depend on running masses in the 
propagators instead of physical masses. The approximation of 
eq.~(\ref{eq:mpolegluinoimp}) does not have this problem, and  
exactly reproduces the direct next-to-leading order width calculation of 
ref.~\cite{Beenakker:1996dw}.

A similar situation holds for the squark width, as shown in the right 
panel of Figure \ref{fig:widths}. In fact, the calculation of 
eq.~(\ref{eq:mpolesquarksnonimp}) gives slightly negative (and therefore 
unphysical) values for the width in a narrow range on either side of the 
physical threshold. This is because the two-loop pole mass contribution to 
the width overcompensates for the one-loop contribution, which is strictly 
positive for all $M_{\tilde Q} \geq 0.969 M_{\tilde g}$. Again, the 
two-loop calculation of eq.~(\ref{eq:mpolesquarksimp}) gets the kinematics 
correct, and precisely reproduces the next-to-leading order calculation of 
ref.~\cite{Beenakker:1996dw}.

I next turn to the effect of the partial three-loop contributions, derived 
in section \ref{sec:threeloopgluino}, on the gluino pole mass. This is 
shown in Figure \ref{fig:threeloop}.%
\begin{figure}[tpb]  
\centering
\mbox{\includegraphics[width=9.1cm]{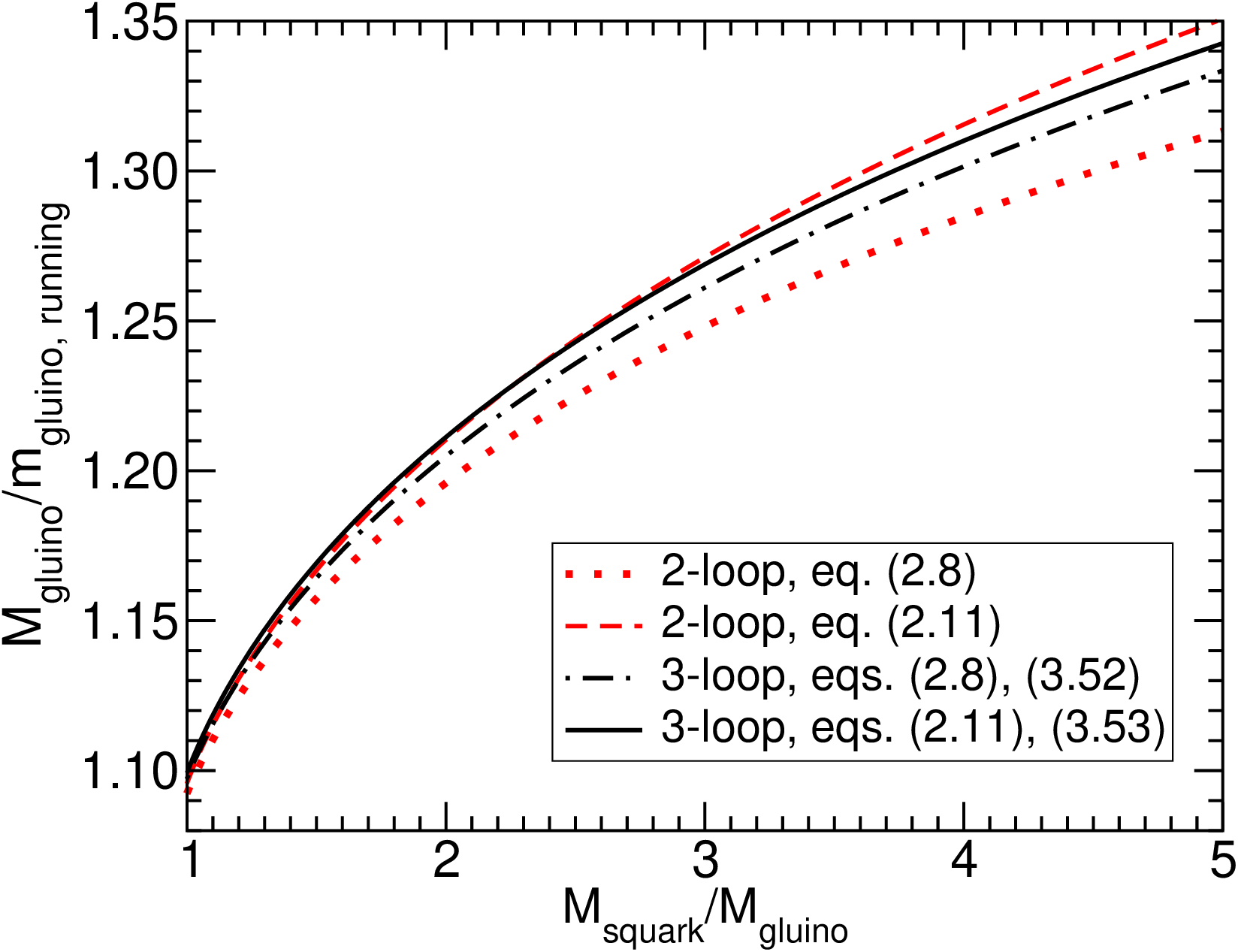}}
\caption{\label{fig:threeloop}
The ratio of the gluino pole mass to the running mass, $M_{\tilde 
g}/m_{\tilde g}(M_{\tilde g})$, as a function of the ratio of the squark 
pole masses to the gluino pole mass, $M_{\tilde Q}/M_{\tilde g}$, as in 
Figure \ref{fig:pole}. The different lines correspond to different 
two-loop and partial three-loop approximations, as defined in the text.}
\end{figure}
Strictly speaking, the three-loop calculations given here are only valid 
in the formal limit $M_{\tilde Q}^2 \gg M_{\tilde g}^2$, but in any case 
the applied correction is small for squark masses just above the gluino 
mass, so I have taken the liberty of showing the entire range $M_{\tilde 
Q} > M_{\tilde g}$. The two three-loop 
approximations are much closer to each other than the corresponding 
two-loop approximations, just as one might have hoped. They differ by less 
than 1\% even for $M_{\tilde Q}/M_{\tilde g} = 5.$ It is also noteworthy 
that both three-loop results are closer to the two-loop approximation of 
eq.~(\ref{eq:mpolegluinoimp}) than they are to 
eq.~(\ref{eq:mpolegluinononimp}), providing some circumstantial evidence 
for the superiority of eq.~(\ref{eq:mpolegluinoimp}).

Finally, consider the renormalization group scale dependence of the 
calculated relationship between the pole mass and the running mass of the 
gluino. Numerical results are shown in Figure \ref{fig:scale}, for three 
ratios $M_{\tilde g}/M_{\tilde Q} = 0.9$, $1.5$, and $3$.%
\begin{figure}[t]  
\centering
\mbox{\includegraphics[width=8.8cm]{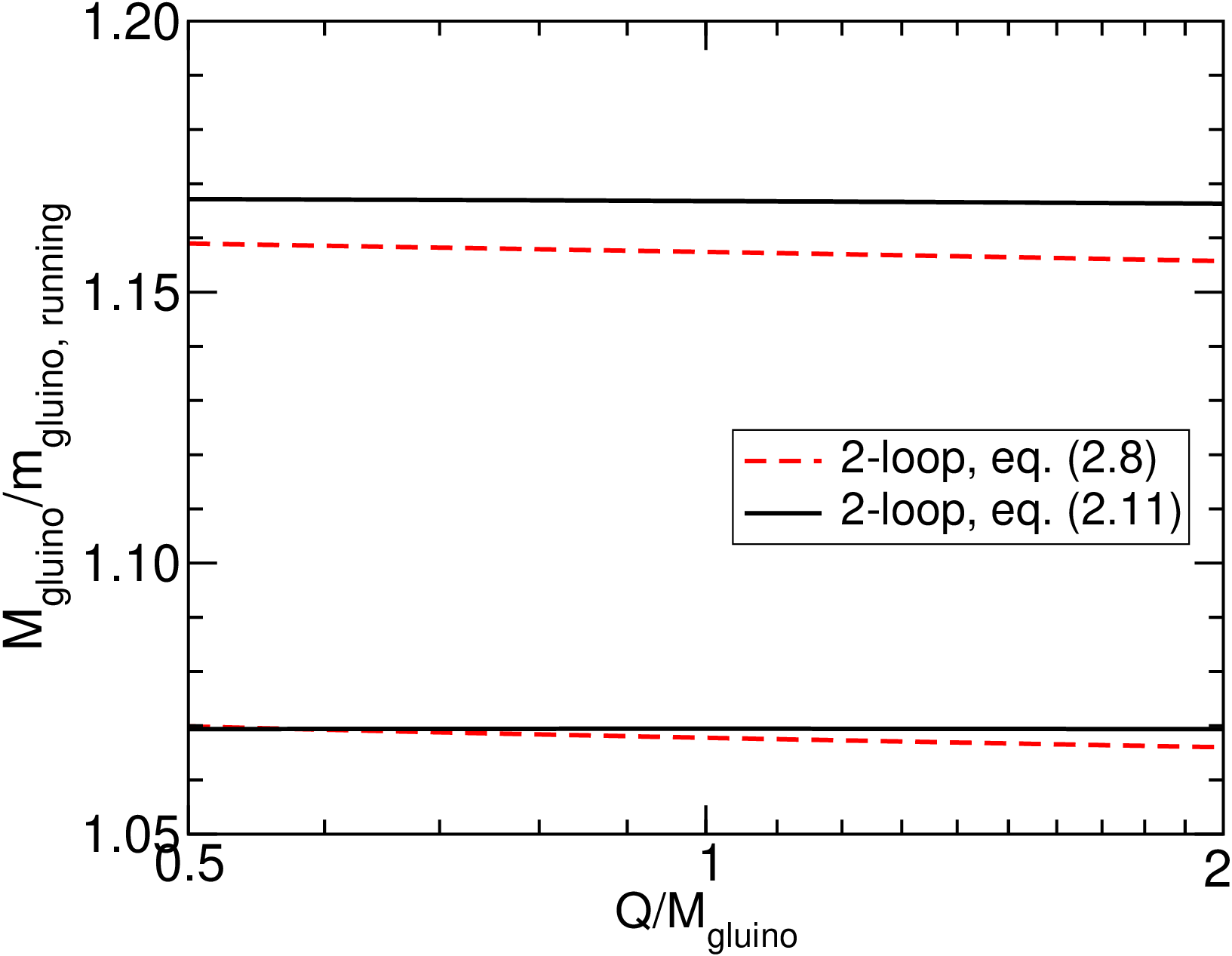}
~
\includegraphics[width=8.8cm]{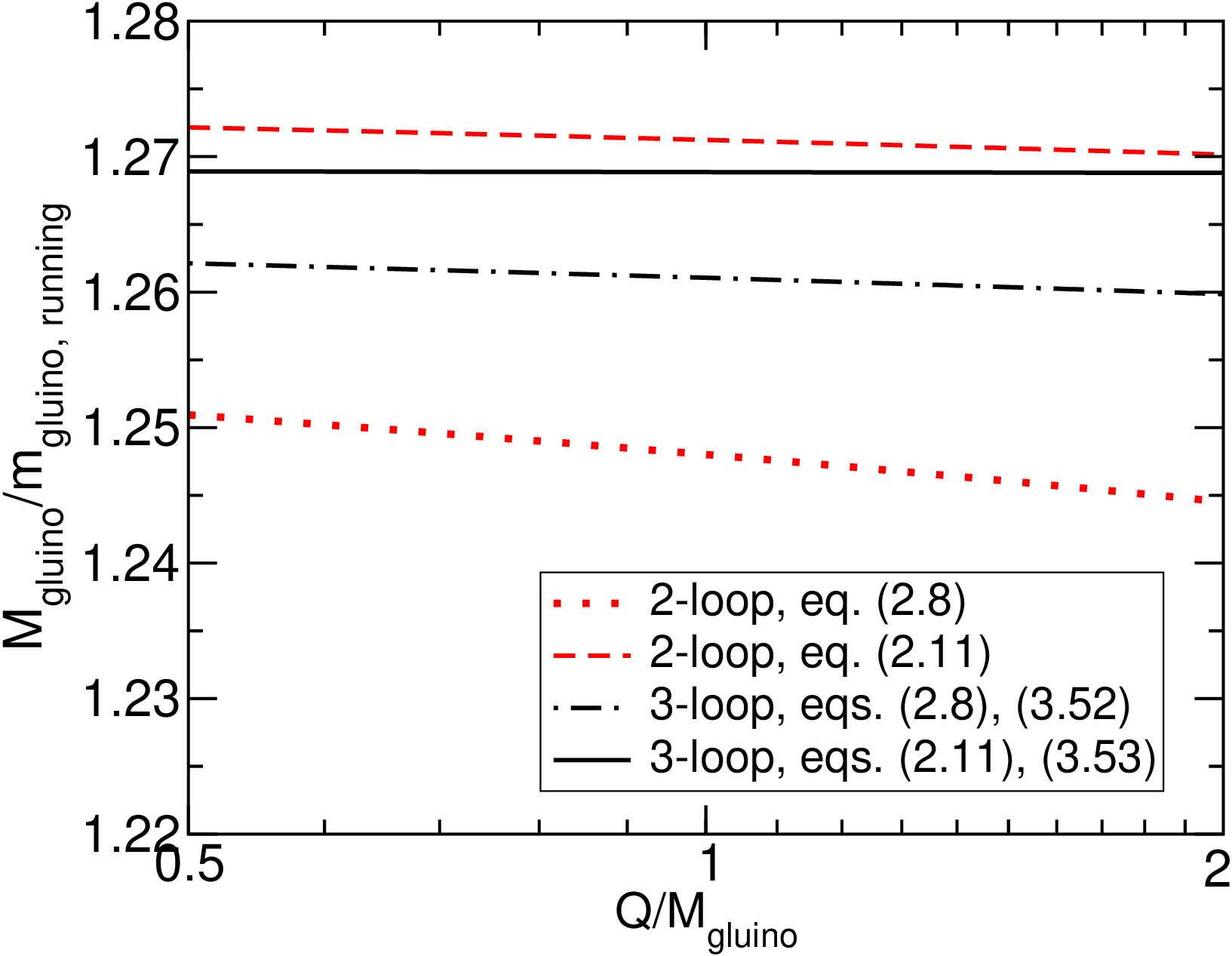}}
\caption{\label{fig:scale}
Renormalization-group scale ($Q$) dependence of the ratio of the gluino 
pole mass to the running mass, $M_{\tilde g}/m_{\tilde g}(M_{\tilde g})$. 
The left panel shows the two-loop approximations of 
eqs.~(\ref{eq:mpolegluinononimp}) and (\ref{eq:mpolegluinoimp}), for 
$M_{\tilde g}/M_{\tilde Q} = 0.9$ (lower pair of lines) and $M_{\tilde 
g}/M_{\tilde Q} = 1.5$ (upper pair of lines). The right panel shows the 
same information, and also the three-loop approximate contributions of 
eqs.~(\ref{eq:gluinothreenonimpnum}) and (\ref{eq:gluinothreeimpnum}), for 
$M_{\tilde g}/M_{\tilde Q} = 3$.}
\end{figure}
In each case, the ratio of the pole mass $M_{\tilde g}$ to the running 
mass evaluated at the pole mass, $m_{\tilde g}(M_{\tilde g})$ is computed 
for a fixed model in terms of the renormalization scale $Q$ at which the 
calculation of the pole mass is performed. The renormalization 
group equations (\ref{eq:betahthree})-(\ref{eq:endfullRGE}) 
are used to run the running parameters between 
different values of $Q$. Comparing the two-loop results, the approximation 
of eq.~(\ref{eq:mpolegluinoimp}) is slightly more stable than that found 
using eq.~(\ref{eq:mpolegluinononimp}), although both are quite acceptably 
scale-invariant for $M_{\tilde g}/M_{\tilde Q} = 0.9$ and $1.5$. In the 
case of $M_{\tilde g}/M_{\tilde Q} = 3$, shown in the right panel, I also 
include the three-loop contributions
of eqs.~(\ref{eq:gluinothreenonimpnum}) and (\ref{eq:gluinothreeimpnum}). 
They exhibit a still further improved scale dependence; 
this is encouraging but cannot be counted as a surprising 
triumph, since the explicit $Q$ 
dependence of the three-loop contribution to the pole mass came from 
nothing other than the three-loop beta functions. 
\end{widetext}

\section{C\lowercase{onclusion}}
\setcounter{equation}{0}
\setcounter{footnote}{1}

In this paper, I have argued in favor of a reformulation of the two-loop 
approximation between pole and running squared masses. As an improvement 
over eq.~(\ref{eq:mpolegen}), equation (\ref{eq:mpolegenimp}) has general 
applicability. It was applied here to the specific case of gluinos and 
squarks in the SUSYQCD sector of the MSSM in 
eqs.~(\ref{eq:mpolesquarksimp}) and (\ref{eq:mpolegluinoimp}). I also used 
the method of effective field theory to obtain a partial three-loop 
approximation to the gluino pole mass, when squarks are heavier. The 
agreement between the three-loop gluino pole mass results and the two-loop 
approximation of eq.~(\ref{eq:mpolegluinoimp}) provides evidence that the 
method of expanding running masses about the real part of pole masses in 
the loop corrections provides better accuracy. Another piece of evidence 
in favor of this conjecture is the agreement of the imaginary part of the 
pole mass with a direct calculation of the width. Also, the improved 
renormalization scale dependence is at least consistent with it. If the 
LHC discovers strongly interacting superpartners, then the quest to 
decipher the organizing principle behind supersymmetry breaking should 
eventually benefit from the improved results presented here, as well as 
similar applications to the rest of the sparticle spectrum.

\section*{A\lowercase{ppendix:} T\lowercase{hree-loop results 
for gauge theories with fermions in arbitrary representations}}
\renewcommand{\theequation}{A.\arabic{equation}}
\setcounter{equation}{0}
\setcounter{footnote}{1}

In this Appendix, I compile the following results for a general gauge 
theory renormalized in the $\MSbar$ scheme with fermions in arbitrary 
representations and no scalar fields:
\begin{itemize}
\item the three-loop beta functions for gauge couplings, 
\item the three-loop beta functions for fermion masses, and
\item the three-loop relation between the pole and running $\MSbar$ 
masses, in the limit that there is only one non-vanishing fermion mass
parameter.
\end{itemize}
Each of these results has appeared before in the case of theories that are 
QCD-like (containing a single gauge group and a single type of fermion 
representation). However, there are non-trivial ambiguities in inferring 
the three-loop results for general theories from the published literature. 
The purpose here is to resolve these ambiguities for use in the main text 
of the present paper and for future reference.

To set notation, consider a theory with a gauge group $G$ which is the 
product of one or more simple or $U(1)$ gauge groups $G_a$, each with a 
distinct $\MSbar$ gauge coupling $g_a$. Results will be written in terms 
of the combinations
\beq
h_a \equiv g_a^2/16\pi^2 .
\eeq
Suppose further that the two-component Weyl fermions of the theory 
transform in possibly distinct representations of the gauge group labelled 
by $R$. Each Dirac (Majorana) fermion consists of two (one) such Weyl 
fermions, so this entails no loss of generality. The quadratic Casimir 
invariants of the adjoint and fermion representations of each gauge group 
are written as $C_a(A)$ and $C_a(R)$, respectively. The normalization is 
such that $C_a(A) = N$ for $SU(N)$, and $C_a(R) = (N^2 - 1)/2N$ when $R$ 
is a fundamental representation of $SU(N)$. The Dynkin index of each 
representation $R$ is written as $I_a(R)$, in a normalization such that 
the fundamental representation of $SU(N)$ has index $1/2$ for a Weyl 
fermion. I also define the invariants:
\beq
S_a &=& \sum_R I_a(R),\\
S_{ab} &=& \sum_R I_a(R) C_b(R), \\
S_{abc} &=& \sum_R I_a(R) C_b(R) C_c(R).
\eeq
Note that a Dirac fermion contributes twice to each of these sums. For 
example, a Dirac fermion in a fundamental representation of $SU(N)$ will 
contribute 1 to $S_a$, and a Dirac fermion with charge $q$ under a $U(1)$
gauge group will contribute $2q^2$ to the corresponding $S_a$.

The three-loop beta function for each of the $\MSbar$ gauge couplings is:
\beq
\beta_{h_a} 
\>\equiv\> Q\frac{d}{dQ} h_a
\>=\> 
\beta_{h_a}^{(1)} +\beta_{h_a}^{(2)} +\beta_{h_a}^{(3)} + \ldots
\eeq
where the terms in the loop expansion are:
\beq
\beta_{h_a}^{(1)} &=& h_a^2 [ -22 C_a(A) + 4 S_a ]/3 ,
\eeq
\begin{widetext}
\beq
\beta_{h_a}^{(2)} &=& 
h_a^3 C_a(A) \bigl [ -68 C_a(A) + 20 S_{a} \bigr ]/3 
+ 4 h_a^2 h_b S_{ab},
\\
\beta_{h_a}^{(3)} &=& h_a^4 C_a(A) \bigl \lbrace 
-2857 [C_a(A)]^2 + 1415 C_a(A) S_a - 79 [S_a]^2 \bigr \rbrace/27
\nonumber \\ &&
+ h_a^3 h_b 8 C_a(A) S_{ab} + 
h_a^2 h_b^2 S_{ab} \bigl [ 133 C_b(A) - 22 S_b\bigr ]/9
- 2 h_a^2 h_b h_c S_{abc}
,
\eeq
\end{widetext}
with the indices $b$ and $c$ implicitly summed over in each term where 
they appear. The special case of this result for a QCD-like theory with a 
single gauge group component and a single type of fermion representation 
was given in ref.~\cite{Tarasov:1980au}. (The four-loop result has also 
been obtained in QCD-like theories, in \cite{vanRitbergen:1997va}). In 
that special case, the three-loop terms proportional to $h_a^3 h_b C_a(A) 
S_{ab}$ and $h_a^2 h_b^2 S_{ab} C_b(A)$ are combined, and the term 
proportional to $h_a^2 h_b^2 S_{ab} S_b$ could in principle combine with a 
term proportional to $h_a^3 h_b S_{ab} S_a$, which is actually absent. 
These two ambiguities have been resolved by considering the special case 
of the electromagnetic coupling beta function with QCD effects included; 
see for example eq.~(54) of \cite{Chetyrkin:1997un} and eqs.~(9)-(10) of 
\cite{Erler:1998sy}.

Next consider the three-loop $\MSbar$ beta function for the mass $m$ of a 
fermion transforming in a representation $F$, given for a QCD-like theory 
in \cite{Tarasov:1982gk}. Chiral symmetry guarantees that if there are 
several such masses, they each run independently. As far as I know, the 
result for a theory with different fermion representations has not been 
given directly in the published literature, but can be inferred by 
considering the results given for individual classes of diagrams in 
\cite{Tarasov:1982gk}. The result is:
\beq
\beta_m 
\>\equiv\> 
Q\frac{d}{dQ} m
\>=\> m \left [ \beta_{m}^{(1)} +\beta_{m}^{(2)} +\beta_{m}^{(3)} 
+ \ldots \right ] ,
\eeq
where the terms in the loop expansion are:
\beq
\beta_{m}^{(1)} &=& -6 h_a C_a(F) ,
\eeq
\begin{widetext}
\beq
\beta_{m}^{(2)} &=& 
h_a^2 C_a(F) \bigl [-97 C_a(A) + 10 S_a \bigr ]/3
- 3 h_a h_b C_a(F) C_b(F) ,
\\
\beta_{m}^{(3)} &=& 
h_a^3 C_a(F) \Bigl \{ -\frac{11413}{54} [C_a(A)]^2 
+ \Bigl [\frac{556}{27} + 48 \zetathree \Bigr ] C_a(A) S_a
+ \frac{70}{27} (S_a)^2 
\Bigr \}
\nonumber \\ &&
+ h_a^2 h_b C_a(F) \Bigl \{
\frac{129}{2} C_a(A) C_b(F) + S_a C_b(F) 
+ [45 - 48 \zetathree] S_{ab} 
 \Bigr \}
-129 h_a h_b h_c C_a(F) C_b(F) C_c(F) .\phantom{xxx}
\eeq
\end{widetext}
with indices $a,b,c$ summed over in terms in which they appear. The terms 
proportional to $C_a(F) S_a C_b(F)$ and $C_a(F) S_{ab}$ are combined in 
the case of quark masses in QCD, and it is this ambiguity that has been 
removed using the results inferred from \cite{Tarasov:1982gk}. (The 
QCD-like case has been extended to four-loop order in 
\cite{fourloopbetaM}.)

Finally, consider the three-loop fermion pole mass. Let the two-component 
fermions consist of massless fermion species with representations labelled 
by $r$, as well as degenerate massive fermion(s) with representation 
labelled by $F$ and a running mass $m(Q)$. This is only technically 
natural if $F$ is irreducible (as for Majorana fermions), or consists of 
an irreducible representation and its conjugate (as for Dirac fermions), 
or if $F$ consists of three or more degenerate copies of a single 
irreducible representation and/or its conjugate (a situation for which I 
know of no examples in proposed extensions of the Standard Model). 
Therefore, it is assumed here that all of the irreducible representations 
labelled by $F$ have the same Casimir invariant $C_a(F)$ and index 
$I_a(F)$. The invariants $S_a$, $S_{ab}$ previously defined are now 
separated into contributions from the massless and massive fermions:
\beq
S_a^L &=& \sum_r I_a(r),
\qquad\qquad\!\!\!\!\!
S_a^H \>=\> \sum_F I_a(F),
\\
S_{ab}^L &=& \sum_r I_a(r) C_b(r),
\qquad\!\!\!\!\!\!
S_{ab}^H = \sum_F I_a(F) C_b(F).
\phantom{xxxxxx}
\eeq
(Again one must remember that the representations are defined for 
two-component fermions, so each Dirac fermion contributes twice to the 
appropriate sums.) Then the fermion pole mass $M$ is related to the 
running $\MSbar$ mass $m$ evaluated at a renormalization scale $Q = M$ by:
\beq
m(M) = M \Bigl [
1 + x^{(1)} + x^{(2)} + x^{(3)}  + \ldots \bigr ] ,
\label{eq:MvRplus}
\eeq
where the loop expansion terms are:
\beq
x^{(1)} &=& -h_a 4 C_a(F) ,
\eeq
\begin{widetext}
\beq
x^{(2)} &=& h_a^2 C_a(F) \bigl \lbrace 
C_a(A) [-1111/24 + 4 \pi^2/3- 4 \pi^2 \ln 2 + 6 \zetathree ]
+ S_a^L [71 + 8 \pi^2]/12 
+ S_a^H [143 - 16 \pi^2]/12 \bigr \rbrace
\nonumber \\ &&
+ h_a h_b C_a(F) C_b(F) [8 \pi^2 \ln 2 - 5 \pi^2 - 12 \zetathree + 7/8]
,
\\
x^{(3)} &=& 64 h_a C_a (F) \Bigl [ 
h_b h_c C_b(F) C_c(F) d^{(3)}_1
+ h_a h_b \bigl\lbrace
C_a(A) C_b(F) d^{(3)}_2
+S_{ab}^L d^{(3)}_{4A}/2
+S_a^L C_b(F) d^{(3)}_{4B}/2
+S_{ab}^H d^{(3)}_{5}/2
\bigr \rbrace\phantom{xx}
\nonumber \\ &&
\!\!\!\!\!\!\!\!\!\!\!
+ h_a^2 \bigl\lbrace
[C_a(A)]^2 d^{(3)}_3
+ C_a(A) S_a^L d^{(3)}_6/2
+ C_a(A) S_a^H d^{(3)}_7/2
+ S_a^L S_a^H d^{(3)}_{8}/4
+ \bigl [S_a^H \bigr ]^2 d^{(3)}_{9}/4
+ \bigl [S_a^L \bigr ]^2 d^{(3)}_{10}/4
\bigr \rbrace
\Bigr ],
\eeq
\end{widetext}
with indices $a,b,c$ summed over wherever they appear. The coefficients 
$d_n^{(3)}$ were found in ref.~\cite{Melnikov:2000qh} for 
$n=1,2,3,5,6,7,8,9,10$ and will not be repeated here. The remaining 
coefficients $d_{4A}^{(3)}$ and $d_{4B}^{(3)}$ were combined into a single 
coefficient $d_{4}^{(3)}$ in that paper, since those terms are 
indistinguishable in the special case of a single type of fermion 
representation. In this paper, I need the generalized result:
\beq
d_{4A}^{(3)} &=& 827/384 + \pi^2/16 - \pi^4/240 - 11 \zetathree/8,
\label{eq:dfoura}
\\
d_{4B}^{(3)} &=& 
\frac{85}{1152} 
+ \frac{95}{144} \pi^2
- \frac{11}{9} \pi^2 \ln 2
+ \frac{11}{3} \zetathree
- \frac{11}{216} \pi^4
\nonumber \\ &&
+ \frac{2}{9} \pi^2 \ln^2 2
+ \frac{1}{9} \ln^4 2
+ \frac{8}{3} \afour.
\label{eq:dfourb}
\eeq 
I obtained $d_{4A}^{(3)}$ by a direct computation of the corresponding 
three-loop diagrams, and then obtained $d_{4B}^{(3)} = d_{4}^{(3)} - 
d_{4A}^{(3)}$ using the result for $d_{4}^{(3)}$ provided in 
ref.~\cite{Melnikov:2000qh}. This was also checked independently using a 
slight modification of the computer code used in 
ref.~\cite{Melnikov:2000qh}, kindly provided by Kirill Melnikov.

In the application of the present paper, the effective theory with squarks 
decoupled consists of an $SU(3)_c$ gauge theory with 6 flavors of 
``massless" Dirac fermion quarks and 1 massive color octet Majorana 
gluino. Therefore, $h_a = \alpha_S/4\pi$ and the relevant group theory 
invariants are:
\beq
&&
\!\!\!\!\!\!
C_a(A) = C_a(F) = 3, 
\\
&&
\!\!\!\!\!\!
S_a^L = 6, \qquad S_{aa}^L = 8, \qquad S_{aaa}^L = 32/3,
\\
&&
\!\!\!\!\!\!
S_a^H = 3, \qquad S_{aa}^H = 9, \qquad S_{aaa}^H = 27,
\\
&&
\!\!\!\!\!\!
S_a = 9, \qquad S_{aa} = 17, \qquad S_{aaa} = 113/3.
\phantom{xxx}
\eeq

{\bf Acknowledgments:} I am grateful to Kirill Melnikov for supplying the 
computer code used in ref.~\cite{Melnikov:2000qh}, which I used to check 
eqs.~(\ref{eq:dfoura}) and (\ref{eq:dfourb}) of the present paper. I
also thank Oleg Tarasov for a communication regarding 
ref.~\cite{Tarasov:1982gk}. This work was supported by the National 
Science Foundation under Grant No.~PHY-0456635.

\end{document}